\documentclass[twocolumn]{autart}    % Enable this line and disable the
\usepackage{natbib}
\usepackage{amsmath,amssymb,bm}  % improve math presentation
\usepackage{graphicx} % takes care of graphic including machinery
\usepackage{color}
\usepackage{algorithm}
\usepackage{algpseudocode}
\usepackage{pgfplots}
\newcommand{\poly}{\bm{\mathcal{C}}}
\newcommand{\indicator}[2]{\mathbb{I}_{\{#1\}}(#2)}
\newcommand{\Tc}[2]{T_{\{#1\}}(#2)}
\newcommand{\Tcn}[2]{T_{\{#1\}}(#2)}

\newcommand{\smallmat}[1]{\left[ \begin{smallmatrix}#1 \end{smallmatrix} \right]}

\newtheorem{definition}{Definition}
\newtheorem{theorem2}{\textbf{Theorem}}{\bf}{\normalfont}

\newtheorem{lemma2}{\textbf{Lemma}}{\bf}{\normalfont}
\newtheorem{remark2}{\textbf{Remark}}{\bf}{\normalfont}

\usepackage{epstopdf}
\usepackage{physics}
\usepackage{amsmath}
\usepackage{tikz}
\usepackage{mathdots}
\usepackage{yhmath}
\usepackage{cancel}
\usepackage{color}
\usepackage{siunitx}
\usepackage{array}
\usepackage{multirow}
\usepackage{amssymb}
\usepackage{gensymb}
\usepackage{tabularx}
\usepackage{booktabs}
\usetikzlibrary{patterns}
\usetikzlibrary{shadows.blur}
\usetikzlibrary{shapes}
\usepackage{subcaption}

\usepackage{soul}

\begin{document}

\newcounter{tempEquationCounter}
\newcounter{thisEquationNumber}
\newenvironment{floatEq}
{\setcounter{thisEquationNumber}{\value{equation}}\addtocounter{equation}{1}% record equation as happened and remember number
\begin{figure*}[!t]% float following equation across columns
\normalsize\setcounter{tempEquationCounter}{\value{equation}}% record current equation number in floated location
\setcounter{equation}{\value{thisEquationNumber}}% use previous equation number
}
{\setcounter{equation}{\value{tempEquationCounter}}% set back to equation number in floated location
\hrulefill\vspace*{4pt}% add a horizontal rule separator
\end{figure*}% end float environment

}
\newenvironment{floatEq2}
{\setcounter{thisEquationNumber}{\value{equation}}\addtocounter{equation}{1}% record equation as happened and remember number
\begin{figure*}[!t]% float following equation across columns
\normalsize\setcounter{tempEquationCounter}{\value{equation}}% record current equation number in floated location
\setcounter{equation}{\value{thisEquationNumber}}% use previous equation number
}
{\setcounter{equation}{\value{tempEquationCounter}}% set back to equation number in floated location
%\hrulefill\vspace*{4pt}% add a horizontal rule separator
\end{figure*}% end float environment

}
\begin{frontmatter}

\title{Computation of Parameter Dependent Robust Invariant Sets for LPV Models with Guaranteed Performance\thanksref{footnoteinfo}} % Title, preferably not more

\thanks[footnoteinfo]{This work is supported by the Vinnova FFI Complex
Control Program, under the grant No. 2015-02309. This paper was not presented at any IFAC meeting. Corresponding author: Ankit Gupta. %Tel. +XXXIX-VI-mmmxxi.
%Fax +XXXIX-VI-mmmxxv.
}

\author[1]{Ankit Gupta}\ead{ankit.gupta@chalmers.se},    % Add the
\author[2]{Manas Mejari}\ead{manas.mejari@supsi.ch},  % e-mail address
\author[1]{Paolo Falcone}\ead{paolo.falcone@chalmers.se},  % (ead) as shown
\author[2]{Dario Piga}\ead{dario.piga@supsi.ch}  % e-mail address

\address[1]{Department of Electrical Engineering, Chalmers University of Technology, Gothenburg, Sweden}  %
\address[2]{IDSIA Dalle Molle
	Institute for Artificial Intelligence, SUPSI-USI, Via Cantonale 2C, 6928 Manno, Switzerland.}  %
                             % keyword list or with the
                                          % help of the Automatica
                                          % keyword wizard

\begin{abstract}                        
%This paper presents an iterative algorithm to compute a robust control invariant (RCI) set for LPV systems. Due to the availability of real-time measurement of the parameters, we allow the RCI set description along with the invariance inducing controller to be parameter-dependent. The considered RCI set and controller description thus lead to parameter-dependent conditions for invariance, which are replaced by sufficient LMI conditions with the help of Polya's relaxation. The LMI conditions are then combined with a novel volume maximization approach in an SDP, which is solved at each iteration of the algorithm to obtain desirably large RCI set and the controller. The algorithm is constructed in a way that a larger RCI set is obtained successively until convergence. The computed invariant sets are always symmetric around the origin by design and have desired representational complexity. In addition to invariance, it is also possible to impose a chosen $\mathcal{H}_2$  and $\mathcal{H}_\infty$ performance level as additional LMI constraints in the SDP. As an outcome, we have shown that the presented algorithm can generate invariant sets larger than the maximal RCI sets, which are incapable of exploiting parameter information. 
This paper presents an iterative algorithm to compute a \emph{Robust Control Invariant} (RCI) set, along with an invariance-inducing control law,  for \emph{Linear Parameter-Varying} (LPV) systems. As  real-time measurements of the scheduling parameters are typically available, we allow the RCI set description and  the invariance-inducing controller to be scheduling parameter dependent. Thus, the considered formulation leads to parameter-dependent conditions for the set invariance, which are replaced by sufficient \emph{Linear Matrix Inequalities} (LMIs)  via Polya's relaxation. These LMI conditions are then combined with a novel volume maximization approach in a \emph{Semidefinite Programming} (SDP) problem, which aims at computing the desirably large RCI set. Besides ensuring invariance, it is also possible to guarantee performance within the RCI set by imposing a chosen quadratic performance level as an additional constraint in the SDP problem. %\sout{The reported numerical example shows that the presented iterative algorithm can generate invariant sets which are larger than the maximal RCI sets computed without exploiting scheduling parameter information.}
Using numerical examples, we show that the presented iterative algorithm can generate RCI sets for large parameter variations where commonly used robust approaches fail.
\end{abstract}

\end{frontmatter}

\section{Introduction}
%Computation of RCI sets has been an active area of research within the control community for several decades. 
RCI set is a set of system states where a feasible control input always exists, which restricts the future states within the set in the presence of disturbances. These sets have become an essential tool for controller synthesis and stability analysis of linear and nonlinear systems \cite{fb,sr10,mf10,jb05}.  %A survey on the existing approaches can be found in the monograph \cite{fb}. 
%To the best of authors’ knowledge,  LPV systems have received relatively less attention. 
%The LPV modeling paradigm is a natural extension of the LTI framework, in which the property of linearity  between input and output signals is preserved in the dynamic relation. However, this relation can change over time according to a measurable (or estimated) time-varying signal, also known as \emph{scheduling parameter}. %$\xi$} (\MMm{Since example is removed from intro, we can remove the sysmbol $\xi$}). 
%In this way, nonlinear and time-varying dynamics can be embedded in the scheduling parameters while retaining the simplicity of LTI model description.
%\MMr{If the scheduling parameters are function of system states and input, then the system is called as quasi-LPV (qLPV)}{}\MMm{line can be moved where qlpv is remarked}.
%Thus, LPV formulations provide a unified framework for the control of linear and many nonlinear systems, see \cite{mj2012}. 

When computing RCI sets for  LPV systems, a common practice is to treat the scheduling parameters as bounded uncertainties \cite{jh20,sm05,ag19b,Nguyen15}.
%Invariant sets play a fundamental role in the control of constrained LPV systems. To list a few  of many important applications, these sets are used for the stability analysis \cite{js95}, in fault detection applications~\cite{tan19} and synthesis of reference governor systems \cite{lc05,eg17} . They are a crucial ingredient in the design of LPV \emph{Model Predictive Control} (MPC) schemes \cite{tb12,jh20} to guarantee persistence feasibility of the underlying optimization problem. %In \cite{nh14}, RCI sets are used to construct an interpolating controller.
Moreover, the invariance inducing control laws are typically assumed to be only state-dependent, without exploiting the observed scheduling parameter information. In this way, the obtained RCI sets can be potentially conservative and, in the worst case,   even empty. %The conservatism can also be understood from the fact that a system which is gain-scheduled stabilizable may not be stabilized robustly \cite{fb07}.
%In this paper, we propose an algorithm to compute a scheduling parameter dependent RCI (PD-RCI) set and its associated gain scheduled control law for polytopic LPV systems. To the best of our knowledge, this work is the first contribution addressing the computation of PD-RCI set. In most of the approaches proposed in the literature CITE, a common practice is to treat the scheduling parameters as bounded uncertainties, for computing the RCI sets, without explicitly taking into account the measured value of the scheduling parameter. Moreover, the computed invariance inducing control laws are assumed to be only state dependent, without exploiting the observed scheduling parameter information. In this way, the obtained RCI sets can be potentially conservative, or in the worst case, they are empty. The advantages of using a parameter dependent RCI set and parameter dependent control laws are motivated as follows:
Thus, to exploit the  information on the scheduling parameters, we propose a new algorithm to compute scheduling parameter-dependent RCI sets and invariance inducing control laws for LPV systems. In this paper, such sets are termed as parameter-dependent RCI (PD-RCI) sets and parameter-dependent control laws (PDCLs), respectively. The advantages of using a PDCL and PD-RCI set are:
%While computing RCI sets for LPV systems, a common practice is to treat parameters as uncertainties. The computed invariance inducing inputs are assumed to be only state dependent, even if parameters are observable. Hence the obtained sets can be potentially conservative, or in the worst case, they are empty. For example, consider the following toy system for which we want to compute the invariant set,
\begin{itemize}
\item \emph{PDCL:} these control laws can stabilize LPV systems %that may not be %robustly stabilizable \cite{fb07}.
that may not be stabilizable by treating the parameters as unknown bounded uncertainties~\cite{fb07}.
%Moreover, our main aim is to propose an algorithm that computes an  RCI set with a desirably large volume. 
Moreover, when computing the RCI sets, keeping PDCL as an optimization variable provides  extra degrees of freedom. We remark that a similar construction was proposed in a robust framework in  \cite{tb10,ag19b,lc19}. 
\item \emph{PD-RCI sets}: Scheduling parameters affect the system's time evolution, and thus the set of states for which invariance can be achieved. Therefore, only considering fixed (or parameter-independent) RCI set description for all scheduling parameters could be restrictive and may lead to conservative (volume-wise) sets. This restrictiveness motivates us to allow the RCI set description to be parameter-dependent.
%The main advantage of having a parameter dependent description is that for a given initial parameter $\xi$, the set of initial states in which invariance is achieved can be made possibly larger than the ones obtained from parameter independent RCI set description. Moreover, the PD-RCI set can be valuable for analysis, since it provides a map between the set of safe initial states and the initial values of scheduling parameters.  % For a given observed current value of the parameter $\xi(t)$, the set $\bm{\mathcal{S}}(\xi(t)$ provides an additional set of values of the safe initial states for which invariance is achieved.  
\end{itemize}
%Generally, the RCI sets only characterize the set of safe initial states for all the possible values of the parameter. In this paper, we present an algorithm to compute an parameter-dependent RCI (PDRCI) set, in which we allow the RCI set description to depend on the parameters (only the initial value, since its future values are unknown). The PDRCI set can be helpful for analysis since it provides a map between the set of safe initial states and the initial values of parameters. Thus if the initial value of the parameter is selectable, then the set of initial states in which invariance is achieved can be possibly larger. Furthermore, having such a map can be advantageous when computing RCI set of a system consisting of many subsystems exchanging states as parameters. 
\vspace{-0.1cm}
This paper presents an iterative algorithm to compute a PD-RCI set of desirably large volume and PDCL for the LPV systems. We  also present a method to compute PD-RCI sets within which a desired quadratic performance can be guaranteed. The representational complexity of the PD-RCI sets can be predefined. The related LMI conditions for invariance are derived by employing Finsler's lemma and Polya's relaxation. These conditions are constructed to ensure invariance for all future (unknown) values of the scheduling parameters. In order to obtain an RCI set with a desirably large volume, we present a volume maximization heuristic based on the theory of Monte-Carlo integration and its convex relaxations.
%The remaining of this paper is structured as follows. In section \ref{sec:problem statement}, we explain the problem we address in the paper. We derive the sufficient parameter-dependent conditions 
%\MM{The paper is organized as follows: TBW}
\vspace{-0.2cm}

{The paper is organized as follows: In Section~\ref{sec:problem statement}, we formalize the problem of computing PD-RCI set and PDCL. Sufficient parameter-dependent conditions for invariance and performance are derived in Section~\ref{sec:suff cond}, and corresponding LMI conditions in Section~\ref{sec:LMI cond}. Using these conditions, an iterative algorithm to compute desirably large RCI sets is proposed in Section~\ref{section:algorithms for rci set computation}. Two case studies are reported in Section~\ref{sec:examples}} %to demonstrate the effectiveness of the proposed algorithm.} %Finally, in Section~\ref{sec:conclusions}, we conclude the paper with some directions for future work. 

\textbf{Notation:}
We use $\mathbb{D}^n_+\in\mathbb{R}^{n\times n}$ to denote  the set of all diagonal matrices with positive diagonal entries.  $I$ and $e_i$ represent the identity matrix and its $i$-th column, and vector of ones is denoted by $\bm{1}$, with dimension  defined by the context. %$\mathbb{S}^n \in \mathbb{R}^{n\times n}$ and $\mathbb{S}^n_+\in \mathbb{R}^{n\times n}$ are, respectively, the set of symmetric and positive definite matrices.  
$X \succ 0\,(\succeq 0)$ denotes a
positive (semi) definite matrix $X$. For compactness, in the text $*$'s will represent matrix's entries that are uniquely identifiable from symmetry, and for some square matrix $X$, $\mathsf{He}(X)=X+X^T$.  We use $X^k$ and $x^k$ to represent a matrix and a vector of appropriate dimension indexed with `$k$'. Let $L(X^k,Y^l,\bar{\Theta},\Theta)$ be a matrix-valued function, where $X^k$ and $Y^{l}$ represent all matrices indexed with `$k$' and `$l$', and  $\bar{\Theta},\Theta$ are some arbitrary matrices. We use $\bm{L}^{k,l}(\bar{\Theta},\Theta) = L(X^k,Y^l,\bar{\Theta},\Theta)$, $\bm{L}^{l,k}(\bar{\Theta},\Theta) = L(X^l,Y^k,\bar{\Theta},\Theta)$ and $\bm{L}^{k,k}(\bar{\Theta},\Theta) = L(X^k,Y^k,\bar{\Theta},\Theta)$. %Given two sets $\bm{\mathcal{X}}$ and $\bm{\mathcal{Y}}$, $\bm{\mathcal{X}}\oplus \bm{\mathcal{Y}}$ represents Minkowski sum of the sets.

\section{Preliminaries}
We recall two existing results which will be used in the paper.
\begin{lemma2}[Finsler's Lemma]\label{lemma: finsler lemma}
 Let $\xi \in \bm{\Xi} \subseteq \mathbb{R}^{N_\xi}$, $\Phi:\bm{\Xi}\rightarrow \mathbb{R}^{n \times n}$ and $\Delta: \bm{\Xi}\rightarrow \mathbb{R}^{m\times n}$. Then the following statements are equivalent~\citep{ji17}:\vspace{-0.2cm}
 \begin{enumerate}
     \item[$i$.] For each $\xi \in \bm{\Xi}$, $y^T\Phi(\xi) y\succ 0,\; \forall \Delta(\xi) y=0,\;y\neq0$.
     \item[$ii$.] For each $\xi \in \bm{\Xi}$, $\exists \Psi \in \mathbb{R}^{n\times m}$ such that \\\hspace*{1cm}$\Phi(\xi)+\Psi(\xi) \Delta(\xi)+\Delta(\xi)^T \Psi(\xi)^T\succ 0$.\vspace{-0.2cm}
 \end{enumerate}
 \end{lemma2}

  \begin{lemma2}[Linearization Lemma]\label{lemma:sucessive linearization lemma} Let $\mathcal{L}\in\mathbb{R}^{m\times n}$ be any arbitrary matrix and  $\mathcal{M}\in\mathbb{R}^{m\times m}$ be a positive definite matrix. The following relation always holds for any arbitrary matrix $\mathcal{Y}\in\mathbb{R}^{m\times n}$~\citep{ag19b}\vspace{-0.2cm}
  \begin{equation}\label{eq:linearization equation}
      \mathcal{L}^T \mathcal{M}^{-1}\mathcal{L}\succeq \mathcal{L}^T\mathcal{Y}+\mathcal{Y}^T\mathcal{L}-\mathcal{Y}^T\mathcal{M}\mathcal{Y}\vspace{-0.2cm}
  \end{equation}
  \end{lemma2}
The result can be easily verified by adding a residual term $(\mathcal{L}-\mathcal{M}\mathcal{Y})^T\mathcal{M}^{-1}(\mathcal{L}-\mathcal{M}\mathcal{Y})$ on the \emph{r.h.s} of \eqref{eq:linearization equation}. This lemma is utilized to resolve the \emph{non-linear} matrix inequalities, i.e., the non-linear term on the l.h.s. of \eqref{eq:linearization equation}    can be replaced with \emph{linear} (in variables $\mathcal{L}, \mathcal{M}$) matrix  term on r.h.s., by appropriate choice of $\mathcal{Y}$.

  \begin{remark2}\label{remark:linearization}[Successive linearization]
  Though $\mathcal{Y}$ can be any arbitrary matrix of compatible dimension, in order to reduce conservatism due to linearization, we suggest successive linearization approach. An appropriate  choice would be to select $\mathcal{Y}=\mathcal{M}^{-1}_0\mathcal{L}_0$, where $\mathcal{M}_0$, $\mathcal{L}_0$ are the values of $\mathcal{M}$, $\mathcal{L}$ obtained from previous iteration. Thus the nonlinearity can be resolved iteratively in which the residual term shrinks with each iteration. Notice that (\ref{eq:linearization equation}) holds with equality if $\mathcal{L}=\mathcal{L}_0$ and $\mathcal{M}=\mathcal{M}_0$, this will be a key property towards proving recursive feasibility of our iterative schemes proposed in this paper. 
 \end{remark2}

%  \MM{\begin{remark2}\label{remark:linearization}[Successive linearization]
% The nonlinearity (\ref{eq:linearization equation}) can be resolved iteratively using successive linearization technique, such that the residual term shrinks with each iteration.  
% In order to reduce the conservatism introduced due to such linearization, an appropriate choice of $\mathcal{Y}$ (instead of any arbitrary $\mathcal{Y}$) would be to select $\mathcal{Y}=\mathcal{M}^{-1}_0\mathcal{L}_0$, where $\mathcal{M}_0$, $\mathcal{L}_0$ are the solutions of  (\ref{eq:linearization equation})  obtained at the previous iteration.
% Notice that (\ref{eq:linearization equation}) holds with equality if $\mathcal{L}=\mathcal{L}_0$ and $\mathcal{M}=\mathcal{M}_0$, this will be a key property towards proving recursive feasibility of the iterative schemes proposed in this paper. 
%  \end{remark2}}

\section{Problem Statement}\label{sec:problem statement}
Let us consider a discrete-time polytopic LPV system: 
\vspace{-0.2cm}
\begin{align}\label{eq:lpv system}
     x(t\!+\!1)\!&=\!\mathcal{A}(\xi(t))x(t)+\mathcal{B}(\xi(t))u(t)+\mathcal{E}(\xi(t))w(t),\\\label{eq:system output equation}
z(t)\!&=\!\mathcal{C}(\xi(t))x(t)+\mathcal{D}(\xi(t))u(t),\vspace{-0.2cm}
\end{align}
where time index $t\in \mathbb{Z}_+$, $x(t) \in \mathbb{R}^{n_x}$ and $z(t) \in \mathbb{R}^{n_z}$ are the current state and the output vectors, $x(t+1)$ is the successor state, and $u(t) \in \mathbb{R}^{n_u}$ and $w(t) \in \mathbb{R}^{n_w}$ are the control and the disturbance input vectors, respectively. The system matrices $\mathcal{A}(\xi(t))$, $\mathcal{B}(\xi(t))$, $\mathcal{C}(\xi(t))$, $\mathcal{D}(\xi(t))$ and $\mathcal{E}(\xi(t))$  depend on the time-varying scheduling parameter $\xi(t)$, which takes value in unit simplex,
\vspace{-0.2cm}
\begin{equation}\label{eq:lpv simplex}
\bm{\Xi}=\left \{ \xi\in \mathbb{R}^{N_{\xi}}:\sum_{k=1}^{N_{\xi}}\xi_k=1, \ \ \xi_k \geq 0 \right \}.\vspace{-0.2cm}
\end{equation}
It is assumed that the current value of $\xi(t)$ is always available. The polytopic system matrices are given by
\vspace{-0.2cm}
\begin{equation}\label{eq:affine lpv representation}
    \begin{bmatrix}
\mathcal{A}(\xi(t))\!\!  &\!\!  \mathcal{B}(\xi(t))\!\!  &\!\!  \mathcal{E}(\xi(t))\\ 
\mathcal{C}(\xi(t))\! \! & \!\! \mathcal{D}(\xi(t))\! &\! 0
\end{bmatrix}\!\! =\!\! \sum_{k=1}^{N_{\xi}}\xi_k(t)\begin{bmatrix}
A^k & B^k & E^k\\ 
C^k & D^k & 0
\end{bmatrix},\vspace{-0.2cm}
\end{equation}
with $A^k, B^k, C^k, D^k, E^k$  real matrices of compatible dimensions. The system is subjected to the following polytopic state/input constraints and bounded disturbance:
\vspace{-0.2cm}
\begin{align}\label{eq: constraints}
\begin{matrix}
\bm{\mathcal{X}}_u&=&\left \{(x,u): H_x x(t)+H_u u(t)\leq \mathbf{1}\right \},\\
%\bm{\mathcal{U}}=&\left \{  u: Gu\leq \mathbf{1}\right \},\\ 
\bm{\mathcal{W}}&=&\hspace{-1cm}\left \{ w: -\mathbf{1}\leq Gw(t)\leq \mathbf{1}\right \}.
\end{matrix}\vspace{-0.2cm}
\end{align}
where $H_x \in \mathbb{R}^{n_h\times n_x}$, $H_u \in \mathbb{R}^{n_h\times n_u}$ and $G \in \mathbb{R}^{n_{g}\times n_w}$ are given matrices. In this paper, we want to compute a 0-symmetric PD-RCI set with a predefined complexity $n_p$ described as
\vspace{-0.2cm}
\begin{equation}\label{eq:invariant set description}
\bm{\mathcal{S}}(\xi(t))\!=\!\left \{  x\in\mathbb{R}^{n_x} : -\bm{1}\!\leq\! \mathcal{P}(\xi(t))W^{-1}x(t) \!\leq\! \bm{1} \right \},\vspace{-0.2cm}
\end{equation} 
where $\mathcal{P}(\xi(t))\triangleq \sum_{k=1}^{N_{\xi}}\xi_k(t)P^k$, $P^k\in\mathbb{R}^{n_p \times n_x}$ and $W \in \mathbb{R}^{n_x\times n_x}$. The presented parameterization of PD-RCI set will be justified when we formalize the problem. Note that, if $P^k = P$ for all $k=1,\ldots,N_{\xi}$, then $\mathcal{P}(\xi(t))=P$, which is similar to the (parameter-independent) RCI set description considered in \cite{ag19a,lc19}. In order to have a non-empty and bounded set $\bm{\mathcal{S}}(\xi(t))$, the matrix $W$ should be invertible and $\mathsf{Rank}(\mathcal{P}(\xi(t)))=n_x,\;\forall \xi \in \bm{\Xi}$. This will be later guaranteed by proper LMI conditions. Furthermore, invariance in the set $\bm{\mathcal{S}}(\xi(t))$ is achieved with a PDCL, which is not known a priori and expressed as\vspace{-0.2cm}
\begin{equation}\label{eq:gain schedule feedback}
    u(t)=\mathcal{K}(\xi(t))x(t),\vspace{-0.2cm}
\end{equation}
where $\mathcal{K}(\xi(t))\triangleq\sum_{k=1}^{N_{\xi}}\xi_k(t)K^k$ and $K^k\in \mathbb{R}^{n_u\times n_x}$. The closed-loop representation of the system \eqref{eq:lpv system} and \eqref{eq:system output equation} with the controller \eqref{eq:gain schedule feedback}  can be written as\vspace{-0.2cm}
\begin{align}\label{eq:closed loop state space dynamics}
    &x(t\!\!+\!\!1)\!\!=\!\!\overbrace{(\!\mathcal{A}(\xi(t))\!+\!\mathcal{B}(\xi(t)\!)\mathcal{K}(\xi(t))}^{\mathcal{A_K}(\xi(t))}x(t)\!\!+\!\!\mathcal{E}(\xi(t))w(t),\\\label{eq:closed loop output equation}
    &z(t)=\overbrace{(\mathcal{C}(\xi(t))\!+\!\mathcal{D}(\xi(t))\mathcal{K}(\xi(t))}^{\mathcal{C_K}(\xi(t))}x(t).\vspace{-0.2cm}
\end{align}
To the best of authors' knowledge, there is no related work which computes the described PD-RCI set. Thus we first formalize the definition of the set by adapting the standard definition of the RCI set to the LPV setting in the sequel.

\begin{definition}
We say a set $\bm{\mathcal{S}}(\xi(t))$ is a PD-RCI set if for any given $\xi(t)\in \bm{\Xi}$ and each $x(t)\in \bm{\mathcal{S}}(\xi(t))$ \vspace{-0.2cm}
\begin{equation}\label{eq:invariance Condn in defn}
\mathcal{A_K}(\xi(t))\bm{\mathcal{S}}(\xi(t))\oplus\mathcal{E}(\xi(t))\bm{\mathcal{W}} \subseteq \bm{\mathcal{S}}(\xi(t+1)), \forall\xi(t\!+\!1)\in \bm{\Xi}\vspace{-0.2cm}
\end{equation}
\end{definition}
Condition \eqref{eq:invariance Condn in defn} should be satisfied for $\forall\xi(t+1)\in \bm{\Xi}$ since $\xi(t+1)$ is unknown at time $t$, which also implies $x(t+1)\in \bigcap_{\forall\xi(t+1)\in\bm{\Xi}} \bm{\mathcal{S}}(\xi(t+1))$.
The computed set $\bm{\mathcal{S}}(\xi(t))$ and the PDCL $\mathcal{K}(\xi(t))$ should obey the system constraints \eqref{eq: constraints}, which implies\vspace{-0.2cm}
\begin{equation}\label{eq:State Constraint in defn}
\bm{\mathcal{S}}(\xi(t))\subseteq\bm{\mathcal{X}}(\xi(t)),\vspace{-0.2cm}
\end{equation}
where $\bm{\mathcal{X}}(\xi(t))\!=\!\left \{  x: (H_x\!+\!H_u\mathcal{K}(\xi(t)))x(t)\leq \mathbf{1}\right \}$.
%Note that,
%\label{rem:classical vs pdrci set}
In  classical formulation,  RCI set description is independent of the scheduling parameters and is  fixed  for all  $\xi \in \bm{\Xi}$. On the other hand, for PD-RCI set \eqref{eq:invariant set description}, different values of initial parameter $\xi(t) \in \bm{\Xi}$ provide different slices\footnote{For a fixed $\bar{\xi} \in \bm{\Xi}$, a \emph{slice}  $\bm{\mathcal{S}}(\bar{\xi})$ is defined as $\bm{\mathcal{S}}(\bar{\xi})= \!\left \{  x\in\mathbb{R}^{n_x} : -\bm{1}\!\leq\! \mathcal{P}(\bar{\xi})W^{-1}x \!\leq\! \bm{1} \right \}$} of the set $\bm{\mathcal{S}}(\xi(t))$ for which the invariance can be achieved. Thus, PD-RCI set provides more flexibility in finding the set of initial states for which invariance can be achieved.

In some applications (e.g., MPC), it may be desirable to have a guaranteed performance within the PD-RCI sets for the closed loop system \eqref{eq:closed loop state space dynamics} and \eqref{eq:closed loop output equation}. For this purpose, we consider quadratic performance constraint\vspace{-0.2cm}
 \begin{equation}\label{eq:H2 cost}
    \sum_{t=0}^{\infty}\left \| z(t) \right \|^2_2\leq \gamma,\;0\leq\gamma<\infty.\vspace{-0.2cm}
\end{equation}
Here \emph{w.l.o.g.}, we assume that performance is measured from time $t=0$. Note that \eqref{eq:H2 cost} can be only satisfied if $w(t) = 0 , \ \forall t\geq0$ (or $w(t)$ eventually becomes zero after certain time). Hence,  we will assume $w(t) = 0$ only when performance constraints are considered.

Our aim  is to compute $\mathcal{P}(\xi(t))$, $W$ and $\mathcal{K}(\xi(t))$, which  define the PD-RCI set \eqref{eq:invariant set description} and the invariance inducing controller \eqref{eq:gain schedule feedback}. We remark that, with $W=I$, computation of $\mathcal{P}(\xi(t))$ and $\mathcal{K}(\xi(t))$ results in a highly non-linear problem.  Indeed, introduction of the matrix $W$ helps overcome the nonlinearity by decomposing the problem into two subproblems described as follows.
%To avoid solving a highly nonlinear problem and for the convenience of our presentation, we divide this computation into two subproblems.  
The first subproblem aims to compute $W$ and $\mathcal{K}(\xi(t))$ for given \emph{parameter-independent} matrix $P$. The second subproblem, aims to compute the parameter-dependent matrix $\mathcal{P}(\xi(t))$ and updated controller $\mathcal{K}(\xi(t))$, for a given matrix $W$ obtained from solving the first subproblem. The two subproblems are formalized as follows. 

\begin{prob}\label{Problem Formulation for invariance condition 1}
For a given matrix $P_{init}\in \mathbb{R}^{n_p \times n_x}$ such that $\mathcal{P}(\xi(t)) = P_{init}$ and the discrete-time system (\ref{eq:lpv system}) subject to constraints (\ref{eq: constraints}), find a matrix $W$ and the control law $\mathcal{K}(\xi(t))$ that satisfies conditions \eqref{eq:invariance Condn in defn}, \eqref{eq:State Constraint in defn} and \eqref{eq:H2 cost} for any arbitrary variation of $\xi(t)\in \bm{\Xi},\forall t\geq 0$.
\end{prob}

\begin{prob}\label{Problem Formulation for invariance condition 2}
For a given matrix $W$ and the discrete-time system (\ref{eq:lpv system}) subject to constraints (\ref{eq: constraints}), find the matrix $\mathcal{P}(\xi(t))$ and the control law $\mathcal{K}(\xi(t))$ that satisfies conditions \eqref{eq:invariance Condn in defn}, \eqref{eq:State Constraint in defn} and \eqref{eq:H2 cost} for any arbitrary variation of $\xi(t)\in \bm{\Xi},\forall t\geq 0$.
%\begin{enumerate}
%\item the controlled system in (\ref{eq:lpv system}) satisfies (\ref{eq:invariance Condn in defn});
%\item all elements of the set $\bm{\mathcal{C}}$ in (\ref{eq:invariant set description}) satisfy (\ref{eq:State Constraint in defn}).
%\end{enumerate}
\end{prob}
Observe that by solving \textbf{Problem~\ref{Problem Formulation for invariance condition 1}}, we obtain an RCI set which is \emph{independent} of the parameter $\xi(t)$ since $\mathcal{P}(\xi(t)) = P_{init}$. In order to obtain a PD-RCI set $\bm{\mathcal{S}}(\xi(t))$, we need to solve \textbf{Problem~\ref{Problem Formulation for invariance condition 1}} and \textbf{Problem~\ref{Problem Formulation for invariance condition 2}} sequentially. In both problems, \eqref{eq:H2 cost} is imposed only if performance is desired. Even though we present our formulation in the form of feasibility problems, our final goal is to design algorithms to compute a desirably large PD-RCI set. 
In the next section, we derive matrix inequality conditions for \eqref{eq:invariance Condn in defn}, \eqref{eq:State Constraint in defn} and \eqref{eq:H2 cost}. These conditions will be later used to obtain  LMI conditions which solve  \textbf{Problem~\ref{Problem Formulation for invariance condition 1}} and \textbf{Problem~\ref{Problem Formulation for invariance condition 2}}. 

\section{Sufficient Parameter Dependent Conditions for Invariance and Performance}\label{sec:suff cond}
For brevity, we will  suppress the time dependent representation of the considered signals and use superscript `$^+$' to indicate successor of $x(t)$ and $\xi(t)$. The arguments of the matrices $\mathcal{A_K}(\xi)$, ${\mathcal{E}}(\xi)$, $\mathcal{C_K}(\xi)$, $\mathcal{P}(\xi)$ and the set $\bm{\mathcal{S}}(\xi)$ will be suppressed and recalled whenever necessary. %Besides this, any new parameter dependent matrices will be we use calligraphic variables to denote any parameter dependent variables. 
%We are now ready to derive sufficient conditions for invariance and the performance constraints.
\subsection{Parameter dependent conditions for invariance and system constrains}
From \eqref{eq:invariant set description} and \eqref{eq:invariance Condn in defn}, a set $\bm{\mathcal{S}}(\xi)$ is invariant, if for a given $\xi \in\bm{\Xi}$ and for each $x\in \bm{\mathcal{S}}(\xi)$, for $i=1,\ldots,n_p$ \vspace{-0.2cm}
\begin{multline}\label{eq: necessary and sufficient condition for invariance}
  \!\! (1 \!-\! (e_i^{T}\mathcal{P}(\xi^+)W^{-1}x^{+})^2)\!\geq \! 0, \forall (w,\xi^+) \! \in \!(\bm{\mathcal{W}},\bm{\Xi}),%\\i=1,\ldots,n_p.\vspace{-0.2cm}
\end{multline}
Using S-procedure (\cite{ip07}), \eqref{eq: constraints} and \eqref{eq:invariant set description}, we can rewrite condition \eqref{eq: necessary and sufficient condition for invariance} as, 
\vspace{-0.3cm}
\begin{multline} \label{eq: sufficient condition for invariance}
\phi_{i}(1 - (e_i^{T}\mathcal{P}(\xi^+)W^{-1}x^{+})^2) \geq \! \\ (\bm{1}\!-\!\mathcal{P}W^{-1}x)^T\Lambda_i (\bm{1}\!+\!\mathcal{P}W^{-1}x)\!+\!(\bm{1}\!-\!Gw)^T\Gamma_i (\bm{1}\!+\!Gw),\\\forall (w,\xi^+)\in (\bm{\mathcal{W}},\bm{\Xi}),i=1,\ldots,n_p,\vspace{-0.2cm}
\end{multline}
where $\phi_{i}\in\mathbb{R}_+$, $\Lambda_i\in \mathbb{D}_+^{n_p}$ and $\Gamma_i\in\mathbb{D}_+^{n_g}$. The vector $x^+$ in \eqref{eq: sufficient condition for invariance} should satisfy \eqref{eq:closed loop state space dynamics}, hence \eqref{eq: sufficient condition for invariance} can be written as
\vspace{-0.2cm}
\begin{multline}\label{eq:invariant condition in finsler form}
    \chi_1^T\begin{bmatrix}
r_i& 0 & 0 & 0\\ 
 0 & W^{-T}\mathcal{P}^T \Lambda_i \mathcal{P}W^{-1} & 0 & 0\\ 
 0 & 0 &  G^T \Gamma_i G & 0\\
 0 & 0 &  0 & -p_i
\end{bmatrix}{\chi_1} \succeq 0,\;\\ \forall\begin{bmatrix}
0&-\mathcal{A_K}&-\mathcal{E}& I
\end{bmatrix}\chi_1=0,
\end{multline}
where $\chi_1=\begin{bmatrix}1&x^T& w^T & (x^+)^T\end{bmatrix}^T$, $r_i=\phi_i-\bm{1}^T\Lambda_i \bm{1}-\bm{1}^T\Gamma_i \bm{1}$ and $p_i= W^{-T}\mathcal{P}^T(\xi^+) e_i \phi_ie_{i}^T\mathcal{P}(\xi^+)W^{-1}$. We will utilize \textbf{Lemma~\ref{lemma: finsler lemma}}
%\textbf{Finsler lemma} (\cite{lmibook}) 
to derive sufficient condition for \eqref{eq:invariant condition in finsler form}. In particular, by choosing $\Psi_i(\xi)  = \begin{bmatrix}
0 & 0 & 0 &\mathcal{V}_{i}(\xi)^{-1}\end{bmatrix}^T$ in \textbf{Lemma~\ref{lemma: finsler lemma}}, 
%(in \textbf{Appendix~\ref{Appendix}}) 
where $\mathcal{V}_{i}(\xi)= \sum_{k=1}^{N_{\xi}}\xi_kV_i^k$,  with $V_i^k\in\mathbb{R}^{n_x \times n_x}$, and by using congruence transform, %\footnote{\MM{Using $\chi_1 = T_c \bar{\chi}_{1}$, where $T_c= \mathrm{diag}(1,W^{T},I, \mathcal{V}^{T}_{i})$}}
we get a sufficient condition for \eqref{eq:invariant condition in finsler form} as follows
\begin{multline}\label{eq: nonlinear matrix inequality sufficient conditions for invariance prior}
\begin{bmatrix}
r_i& 0 & 0 & 0\\ 
 0 & \mathcal{P}^T \Lambda_i \mathcal{P} & 0 & \mathcal{A}_{\bar{\mathcal{K}}}^T\\ 
 0 & 0 &  G^T \Gamma_i G & \mathcal{E}^T\\
 0 & * &  * &\mathsf{He}(\mathcal{V}_i) -\mathcal{V}_{i}^{T}p_i\mathcal{V}_{i}
\end{bmatrix}\succ 0,\forall \xi^+\in \bm{\Xi},\;\\i=1,\ldots,n_p,
\end{multline}
where $\mathcal{A}_{\bar{\mathcal{K}}}=\mathcal{A_K}W$ and  $\bar{\mathcal{K}}(\xi)=\mathcal{K}(\xi)W \triangleq\sum_{k=1}^{N_\xi} \xi_k\bar{K}^k$. With the intention to resolve the nonlinearity in the $(4,4)$-block of (\ref{eq: nonlinear matrix inequality sufficient conditions for invariance prior}), we now introduce a positive-definite matrix variable $X_{i}$ that satisfies
\begin{equation}\label{Xi}
   X_{i}^{-1}-p_i\succ 0.
\end{equation}
Thus, from \eqref{eq: nonlinear matrix inequality sufficient conditions for invariance prior} and \eqref{Xi}, we obtain a sufficient parameter dependent matrix inequality condition for \eqref{eq:invariance Condn in defn} as
\begin{subequations}\label{eq: nonlinear matrix inequality sufficient conditions for invariance}
\begin{align}\label{eq: nonlinear matrix inequality sufficient conditions for invariance 1}
\begin{bmatrix}
W^{T}X_{i}^{-1}W & *\\
\phi_{i}e_{i}^{T}\mathcal{P}(\xi^+) & \phi_{i}
\end{bmatrix}&\succ 0,\\\label{eq: nonlinear matrix inequality sufficient conditions for invariance 2}
\begin{bmatrix}
r_i& 0 & 0 & 0\\ 
 0 & \mathcal{P}^T \Lambda_i \mathcal{P} & 0 & \mathcal{A}_{\bar{\mathcal{K}}}^T\\ 
 0 & 0 &  G^T \Gamma_i G & \mathcal{E}^T\\
 0 & * &  * &\mathsf{He}(\mathcal{V}_i) -\mathcal{V}_{i}^{T}X_i^{-1}\mathcal{V}_{i}
\end{bmatrix}&\succ 0,\\\nonumber\forall \xi^+\in \bm{\Xi},i=1,\ldots,n_p.
\end{align}
\end{subequations}
In the next lemma, we present sufficient parameter dependent conditions for the invariance \eqref{eq:invariance Condn in defn} and system constraints \eqref{eq:State Constraint in defn} . %We would refer readers to \textbf{Lemma~\ref{lemma:sucessive linearization lemma}}, which will be utilized in the derivation. %Furthermore, we introduce some new notations to isolate parameters from the matrices and compactness of our presentation. 
%\MM{It is better to either move it in notations or here create environment notation with a tag}
%Let $L(X^k,Y^l,\bar{\Theta},\Theta)$ be some function of matrices, where $X^k$ and $Y^{l}$ represent all matrices indexed '$k$' and '$l$'. For compactness, we use $\bm{L}^{k,l}(\bar{\Theta},\Theta) = L(X^k,Y^l,\bar{\Theta},\Theta)$, $\bm{L}^{l,k}(\bar{\Theta},\Theta) = L(X^l,Y^k,\bar{\Theta},\Theta)$ and $\bm{L}^{k,k}(\bar{\Theta},\Theta) = L(X^k,Y^k,\bar{\Theta},\Theta)$.
\begin{lemma2}\label{lemma: parameter dependent invariance conditon} 
For some arbitrary matrices $Y_i\in \mathbb{R}^{n_x \times n_x}$, $\bar{\Lambda}_i\in \mathbb{D}_+^{n_p}$, $i=1,\ldots,n_p$, $\bar{\Pi}_j\in \mathbb{D}_+^{n_p}$, $j=1,\ldots,n_h$ and $P^k_0\in \mathbb{R}^{n_p \times n_x}$, $k=1,\ldots,N_\xi$, if there exist matrices $W$, $P^k$, $\bar{K}^k$, $V_i^k$, $X_i$, diagonal semi-definite matrices $\Lambda_i$, $\Gamma_i$, $\Pi_j$ and scalar $\phi_i>0$ satisfying conditions  \eqref{eq:parameter dependent invariance condition 1},\eqref{eq:parameter dependent invariance condition 2}, \eqref{eq:parameter dependent invariance condition 3} and \eqref{eq:parameter dependent system Constraints} reported below, then a PD-RCI set can be obtained as in (\ref{eq:invariant set description}) and the PDCL as $\mathcal{K}(\xi)=\bar{\mathcal{K}}(\xi)W^{-1}$:
\begin{subequations}\label{eq:parameter dependent invariance condition}
\begin{align}
\begin{bmatrix}\label{eq:parameter dependent invariance condition 1}
    W^{T}Y_i+Y_i^{T}W-Y_i^TX_{i}Y_i& *\\
    \phi_{i}e_{i}^{T}\mathcal{P}(\xi^+) & \phi_{i}
\end{bmatrix}\succ 0,\\\label{eq:parameter dependent invariance condition 2}
\phi_i-\bm{1}^T\Lambda_i\bm{1}-\bm{1}^T\Gamma_i \bm{1}\succ 0,
\end{align}\vspace{-1cm}
\begin{multline}\label{eq:parameter dependent invariance condition 3}
    \! \sum_{k=1}^{N_{\xi}}\xi_k^2 \bm{M}_{i}^{k,k}(\bar{\Lambda}_i,\Lambda_i)+ \\ \! \!\sum_{k=1}^{N_{\xi}-1}\!\!\sum_{l=k+1}^{N_{\xi}}\xi_k \xi_l(\bm{M}_{i}^{k,l}(\bar{\Lambda}_i,\Lambda_i)\!+\!\bm{M}_{i}^{l,k}(\bar{\Lambda}_i,\Lambda_i))\!\succ\! 0,
\end{multline}\vspace{-1cm}
\end{subequations}
\begin{multline}\label{eq:parameter dependent system Constraints}
\!\sum_{k=1}^{N_{\xi}}\xi_k^2 \bm{R}_{j}^{k,k}(\bar{\Pi}_j,\Pi_j)+ \\
\!\!\sum_{k=1}^{N_{\xi}-1}\!\sum_{l=k+1}^{N_{\xi}}\xi_k \xi_l(\bm{R}_{j}^{k,l}(\bar{\Pi}_j,\Pi_j)\!+\!\bm{R}_{j}^{l,k}(\bar{\Pi}_j,\Pi_j))\!\succeq \! 0,
\end{multline}
\vspace{-0.2cm}
where, 
\begin{equation}\label{eq:Pkl matrix}
  \! \! \bm{P}^{k,l}(\bar{\Lambda}_i,\Lambda_i)\!=\!\mathsf{He}((P^k)^T \bar{\Lambda}_i P^l_0)\!-\!(P^k_0)^T \bar{\Lambda}_i\Lambda_i^{-1}\bar{\Lambda}_i P^l_0,
\end{equation}
\begin{align}\label{eq:Mkl matrix}
\!\!\bm{M}_{i}^{k,l}(\bar{\Lambda}_i,\Lambda_i)\!&=\!\!\begin{bmatrix}
\bm{P}^{k,l}(\bar{\Lambda}_i,\Lambda_i) & * & * & \!*\\ 
0 &  G^T \Gamma_i G & *&\!*\\
A^kW\!+\!B^k\bar{K}^l &  E^k &\mathsf{He}(V_i^k) & \!*\\
0 &  0 & V_i^k &\! X_i
\end{bmatrix},
\end{align}
\begin{align}\label{eq:Rkl matrix}
\bm{R}_{j}^{k,l}(\bar{\Pi}_j,\Pi_j)\!&=\!\begin{bmatrix}
2-\bm{1}^T\Pi_j\bm{1}  & e_j^T(H_x W+H_u\bar{K}^l) \\ 
* & \bm{P}^{k,l}(\bar{\Pi}_j,\Pi_j)
\end{bmatrix}.
\end{align}
\end{lemma2}
%For the proof refer to~\cite{ag21ARXIV}[\textbf{Lemma~3}].
 \begin{pf}
 For some matrix $Y_i$, we obtain (\ref{eq:parameter dependent invariance condition 1}) by applying \textbf{Lemma~\ref{lemma:sucessive linearization lemma}} in $(1,1)$ block of \eqref{eq: nonlinear matrix inequality sufficient conditions for invariance 1}. Next, we consider \eqref{eq: nonlinear matrix inequality sufficient conditions for invariance 2} which can be rewritten as 
 \begin{equation}\label{eq: compressed nonlinear matrix inequality sufficient conditions for invariance}
     \begin{bmatrix}
    r_i&0\\
     0 &\bar{\mathcal{M}}_i 
     \end{bmatrix} \succ0.
 \end{equation}
 The condition \eqref{eq:parameter dependent invariance condition 2} is directly implied from $(1,1)$ block of \eqref{eq: compressed nonlinear matrix inequality sufficient conditions for invariance}, it also implies $\bar{\mathcal{M}}_i\succ 0$. Furthermore, by using \textbf{Lemma~\ref{lemma:sucessive linearization lemma}} again in the $(1,1)$ block of $\bar{\mathcal{M}}_i$ in \eqref{eq: compressed nonlinear matrix inequality sufficient conditions for invariance} with some matrix $\mathcal{Y}_i=\bar{\Lambda}_i \mathcal{P}_{0}(\xi)$, where $\mathcal{P}_{0}(\xi) = \sum_{k=1}^{N_{\xi}}\xi_kP_0^k$, and then by application of Schur complement lemma,  we get
  \begin{equation}\label{eq:parameter dependent matrix inequality condition for invariance}
  \begin{bmatrix}
  \mathsf{He}(\mathcal{P}^T \bar{\Lambda}_i \mathcal{P}_{0})\!-\! \mathcal{P}_{0}^T\bar{\Lambda}_i\Lambda_i^{-1}\bar{\Lambda}_i \mathcal{P}_{0}& 0 & \mathcal{A}_{\bar{\mathcal{K}}}^T&0\\ 
  * &  G^T \Gamma_i G & \mathcal{E}^T&0\\
  * &  * &\mathsf{He}(\mathcal{V}_i) & \mathcal{V}_i^T\\
  * &  * & * & X_i
  \end{bmatrix}\!\!\succ\! 0.
 \end{equation}
 It is straightforward to verify that \eqref{eq:parameter dependent matrix inequality condition for invariance} can be rewritten in polynomial form\footnote{Deriving the polynomials in \eqref{eq:parameter dependent invariance condition 3}, we recall the introduced notation for matrix valued functions $\bm{L}^{k,l}(\bar{\Theta},\Theta) = L(X^k,Y^l,\bar{\Theta},\Theta)$ and the simplex assumption, $\sum_{k=1}^{N_{\xi}}\xi_k =1.$ } as \eqref{eq:parameter dependent invariance condition 3}. 
 Thus a sufficient condition for invariance condition \eqref{eq: nonlinear matrix inequality sufficient conditions for invariance} is given by \eqref{eq:parameter dependent invariance condition}. Additionally, the PD-RCI set $\bm{\mathcal{S}}$ has to satisfy the state and input constraints  \eqref{eq:State Constraint in defn}. By employing S-procedure, it can be expressed as follows
 \begin{equation}\label{eq:state constraint equivalent condition}
     2(1-(e_j^T(H_x+H_u\mathcal{K})x))\!\succeq\! (\bm{1}-\mathcal{P}W^{-1}x)^T\Pi_j(\bm{1}+\mathcal{P}W^{-1}x),
 \end{equation}
 By expressing \eqref{eq:state constraint equivalent condition} in a quadratic form, using congruence transformation and applying \textbf{Lemma~\ref{lemma:sucessive linearization lemma}} with some matrix $\mathcal{Y}_j= \bar{\Pi}_j \mathcal{P}_{0}(\xi)$, the following sufficient condition is obtained \begin{equation}
 \begin{bmatrix}
 2-\bm{1}^T\Pi_j\bm{1}  & e_j^T(H_x W+H_u\bar{\mathcal{K}}(\xi)) \\ 
 * & \mathsf{He}(\mathcal{P}^T \bar{\Pi}_j \mathcal{P}_{0})- (\mathcal{P}_{0})^T\bar{\Pi}_j\Pi_j^{-1}\bar{\Pi}_j \mathcal{P}_{0}
 \end{bmatrix}\succeq 0,
\end{equation}
 which in turn can be equivalently written as \eqref{eq:parameter dependent system Constraints}. Note that the term $\bm{P}^{k,l}(\bar{\Pi}_j,\Pi_j)$ in \eqref{eq:Rkl matrix} has a similar form as \eqref{eq:Pkl matrix}. \hfill{} \qed
 \end{pf}

\begin{remark2}\label{rem:linearization choice}
A feasible solution to inequalities \eqref{eq:parameter dependent invariance condition} and \eqref{eq:parameter dependent system Constraints} for any arbitrary choice of matrices $Y_i$, $\bar{\Lambda}_i$, $\bar{\Pi}_j$ and $P_0^k$ gives a PD-RCI set $\bm{\mathcal{S}}$ and an invariance inducing PDCL $\mathcal{K}$. {From \textbf{Lemma~\ref{lemma:sucessive linearization lemma}}, we know that the ideal choices of these matrices is  $Y_i=X_i^{-1}W$, $\bar{\Lambda}_i = \Lambda_i$, $\bar{\Pi}_j=\Pi_j$ and $P_0^k=P^k$.} However, the mentioned choices do not resolve the nonlinearities in  \eqref{eq:parameter dependent invariance condition} and \eqref{eq:parameter dependent system Constraints}. In Section~\ref{sec:LMI cond}, we will present a systematic way to select these matrices resolving the nonlinearity, which also helps us to reduce the conservatism introduced due to linearization.
\end{remark2}

%Notice that all inequalities, except \eqref{eq:parameter dependent invariance condition 2}, in \textbf{Lemma~\ref{lemma: parameter dependent invariance conditon}} are parameter dependent matrix inequality conditions. In the current form these inequalities are intractable for solving but it will serve as the base to device LMI conditions for invariance in the coming sections. %We will solve them iteratively using successive linearization approach to compute RCI sets. 
\subsection{Parameter dependent performance constraints}
We next derive parameter dependent matrix inequality conditions for performance constraint \eqref{eq:H2 cost}. Since we consider performance for  $w(t)=0,\;\forall t\geq 0$, we can ignore the matrix $\mathcal{E}$ in \eqref{eq:closed loop state space dynamics}.  Now, let $\mathcal{Q}(\xi) = \sum_{k=1}^{N_{\xi}}\xi_k Q^k\succeq0$ with $Q^k\in\mathbb{R}^{n_x\times n_x}$, then the performance constraint \eqref{eq:H2 cost} is satisfied by the closed-loop system \eqref{eq:closed loop state space dynamics} and \eqref{eq:closed loop output equation} within the set $\bm{\mathcal{S}}$  if (\cite{mk96,lc19}):
\begin{subequations}\label{eq:H2 cost necessary and sufficient conditions}
\begin{equation}\label{eq:H2 cost bound lypunov}
    \left \| \mathcal{Q}^{-1/2}x(t) \right \|^2_2\leq \gamma,\; \forall x(t)\in \bm{\mathcal{S}}(\xi),
\end{equation}\vspace{-0.6cm}
\begin{equation}\label{eq:H2 cost decay condition}
    \left \| \mathcal{Q}^{-1/2}(\xi^+)x^+ \right \|^2_2-\left \| \mathcal{Q}^{-1/2}x(t) \right \|^2_2\leq -\left \| z(t) \right \|^2_2.
\end{equation}
\end{subequations}
It is easy to verify that \eqref{eq:H2 cost necessary and sufficient conditions} implies \eqref{eq:H2 cost} by summing both   sides of \eqref{eq:H2 cost decay condition} from $t=0$ to $t=\infty$. In the next lemma, we present parameter dependent sufficient conditions for \eqref{eq:H2 cost bound lypunov} and \eqref{eq:H2 cost decay condition}.

\begin{lemma2}\label{lem:H_2}
For a given $\gamma >0$, and some arbitrary matrices $\bar{\Upsilon}\in \mathbb{D}^{n_p}$ and $P^k_0\in \mathbb{R}^{n_p \times n_x}$, $k=1,\ldots,N_\xi$, the performance constraints \eqref{eq:H2 cost} is fulfilled by the closed-loop system \eqref{eq:closed loop state space dynamics} and \eqref{eq:closed loop output equation} within the set $\bm{\mathcal{S}}(\xi)$, if there exist matrices $W$, $P^k$, $\bar{K}^k$, $Q^k$, {$Z^k$}, $F^k$ and diagonal semi-definite matrix $\Upsilon$ satisfying the following conditions: 
\begin{subequations}\label{eq:parameter dependent H2 cost constraint}
\begin{equation}\label{eq:parameter dependent H2 cost constraint 1}
\sum_{k=1}^{N_{\xi}}\xi_k^2 \bm{N}^{k,k}+\sum_{k=1}^{N_{\xi}-1}\sum_{l=k+1}^{N_{\xi}}\xi_k \xi_l (\bm{N}^{k,l}+\bm{N}^{l,k})\!\succeq\!0.
\end{equation}
\vspace{-0.2cm}
\begin{multline}\label{eq:parameter dependent H2 cost constraint 2}
\sum_{k=1}^{N_{\xi}}\xi_k^2 \bm{L}^{k,k}(\bar{\Upsilon},\Upsilon)\\+\sum_{k=1}^{N_{\xi}-1}\!\!\sum_{l=k+1}^{N_{\xi}}\xi_k \xi_l (\bm{L}^{k,l}(\bar{\Upsilon},\Upsilon)+\bm{L}^{l,k}(\bar{\Upsilon},\Upsilon))\succeq0.
\end{multline}
\end{subequations}
\end{lemma2}
\vspace{-0.2cm}
where, \vspace{-0.5cm}
\begin{align}\nonumber
\bm{N}^{k,l}\!=\!\begin{bmatrix}
\mathsf{He}(W)-Q^k & * & * & * & *\\  
A^kW\!+\!B^k\bar{K}^l & \mathsf{He}(Z^{k}) & * & * & *\\
0     &  Z^k  &  Q^k & * & *\\
C^kW\!+\!D^k\bar{K}^l &  0  &  0 &\mathsf{He}(F^{k}) &*\\
0     &  0  &  0 & F^k & I
\end{bmatrix},
\end{align}
\begin{align}\nonumber
\bm{L}^{k,l}(\bar{\Upsilon},\Upsilon)\!=\!\begin{bmatrix}
\gamma-\bm{1}^T\Upsilon \bm{1} & * & *\\ 
0 & \bm{P}^{k,l}(\bar{\Upsilon},\Upsilon) & *\\
0 & W & Q^k
\end{bmatrix}.
\end{align}
%For the proof refer to~\cite{ag21ARXIV}[\textbf{Lemma~5}].  
 \begin{pf}
 Let $\chi_2 =\begin{bmatrix}
 x(t)^T&  
 x(t+1)^T&
 z(t)^T
 \end{bmatrix}$. Since $x(t+1)$ and $z(t)$ in \eqref{eq:H2 cost decay condition} have  to satisfy \eqref{eq:closed loop state space dynamics} and \eqref{eq:closed loop output equation}, this can be expressed as
 \begin{equation}\label{eq:H2 cost decay equivalent condition}
 \chi_2^T\begin{bmatrix}
 \mathcal{Q}^{-1}& * &*\\  
 0& -\mathcal{Q}^{-1}(\xi^+)&0\\
 0& 0 & -I
 \end{bmatrix}\chi_2\succeq 0,
 \forall\begin{bmatrix}
 -\mathcal{A_K} & I & 0\\ 
 -\mathcal{C_K} & 0 & I
 \end{bmatrix} \chi_2=0.
 \end{equation}
 Let $\mathcal{Z}= \sum_{k=1}^{N_{\xi}}\xi_k Z^k$ and $\mathcal{F}= \sum_{k=1}^{N_{\xi}}\xi_k F^k$, where $Z^k\in \mathbb{R}^{n_x \times n_x}$ and $F^k\in \mathbb{R}^{n_z \times n_z}$. Choosing $\Psi=\begin{bmatrix}
 0  & \mathcal{Z}^{-1}(\xi) & 0\\ 
 0  &  0 & \mathcal{F}^{-1}(\xi)
 \end{bmatrix}^T$ in \textbf{Lemma~\ref{lemma: finsler lemma}}, and using congruence transform followed by Schur complement, a sufficient condition for \eqref{eq:H2 cost decay equivalent condition} can be expressed as 
 \begin{equation}\label{eq:parameter dependent matrix inequality condition for H2 cost}
 \begin{bmatrix}
 W^T\mathcal{Q}^{-1}W & * & * & * & *\\  
 \mathcal{A}_{\bar{\mathcal{K}}} & \mathsf{He}(\mathcal{Z}) & * & * & *\\
 0     &  \mathcal{Z}  &  \mathcal{Q}(\xi^+) & * & *\\
 \mathcal{C}_{\bar{\mathcal{K}}} &  0  &  0 &\mathsf{He}(\mathcal{F}) & *\\
 0   &  0  &  0 &  \mathcal{F} & I
 \end{bmatrix}\succeq0.
 \end{equation}
 The nonlinearity in the $(1,1)$-block of \eqref{eq:parameter dependent matrix inequality condition for H2 cost} can be resolved by application of \textbf{Lemma~\ref{lemma:sucessive linearization lemma}}. To avoid inverse parameter dependent term (see, \textbf{Remark~\ref{remark:linearization}}), we select $\mathcal{Y}=I$ while linearizing \eqref{eq:parameter dependent matrix inequality condition for H2 cost}. As before, the obtained inequality condition can be thus written in a polynomial form \eqref{eq:parameter dependent H2 cost constraint 1}. 
 Lastly for \eqref{eq:H2 cost bound lypunov}, an equivalent condition obtained by employing S-procedure is 
 \begin{equation}\label{eq:H2 performance set constraints}
     \gamma-x(t)\mathcal{Q}^{-1}x(t)\!\succeq\! (\bm{1}-\mathcal{P}W^{-1}x(t))^{T}\Upsilon (\bm{1}+\mathcal{P}W^{-1}x(t)).
 \end{equation}
 Expressing \eqref{eq:H2 performance set constraints} in a quadratic form and by applying \textbf{Lemma~\ref{lemma:sucessive linearization lemma}} with $\mathcal{Y}=\bar{\Upsilon} \mathcal{P}_{0}(\xi)$, a  parameter dependent matrix inequality condition can be written as \eqref{eq:parameter dependent H2 cost constraint 2}.\hfill{} \qed
 \end{pf}
Notice that the performance constraints \eqref{eq:parameter dependent H2 cost constraint 2} depend on matrices $\bar{\Upsilon}$ and $P_0^k$, and their ideal choices are $\Upsilon$ and $P^k$, respectively. We will present systematic choices of these matrices in the next section.
To summarize, in this section, we have obtained parameter dependent matrix inequality conditions for invariance \eqref{eq:invariance Condn in defn}, system constraints \eqref{eq:State Constraint in defn}, and performance constraints \eqref{eq:H2 cost} which are given by \eqref{eq:parameter dependent invariance condition}, \eqref{eq:parameter dependent system Constraints} (\textbf{Lemma \ref{lemma: parameter dependent invariance conditon}}), and \eqref{eq:parameter dependent H2 cost constraint}  (\textbf{Lemma \ref{lem:H_2}}), respectively. The parameter dependent conditions are linear if $\bm{P}^{k,l}$ is linear. Assuming $P_0^k$ is known, the linearity of the matrix $\bm{P}^{k,l}$ in turn depends on the matrices $\bar{\Lambda}_i$ (and $\bar{\Pi}_j$, $\bar{\Upsilon}$) and $P^k$. Resolving the nonlinearity in $\bm{P}^{k,l}$ was one of the main  motivating factors behind the presented formulation of \textbf{Problem~\ref{Problem Formulation for invariance condition 1}} and \textbf{Problem~\ref{Problem Formulation for invariance condition 2}}. %Furthermore, the matrix inequality conditions are parameter dependent,  hence solving them in the current form can be intractable. Nevertheless, since $\xi_k \in \bm{\Xi}$ are positive, it is easy to verify that a sufficient condition for \eqref{eq:parameter dependent invariance condition 3} would be $\bm{M}_{i}^{k,k}(\bar{\Lambda}_i,\Lambda_i)\succeq0, \; k=1,\ldots,N_{\xi}$ (necessary conditions) and $\bm{M}_{i}^{k,l}(\bar{\Lambda}_i,\Lambda_i)+\bm{M}_{i}^{l,k}(\bar{\Lambda}_i,\Lambda_i)\succeq 0,\; k=1,\ldots,N_{\xi}-1, l=k+1,\ldots,N_{\xi}$ (sufficient conditions). Similar matrix inequality condition can be also obtained for system and performance constraints. From \cite{or05,cs05}, it is known that these sufficient conditions can be conservative. By applying Pólya relaxation, the conservatism can be reduced at the cost of an increased number of total LMI conditions depending upon the choice of the order of relaxation. %For example, let the number of system vertices $N_{\xi}= 2$, now if we select $d=0$ in (\ref{eq:LMI invariance condition 3 for case 1}), we get the same conditions as discussed earlier in this remark i.e. $M_i^{kk}\succeq0, \; k=1,2$ and $M_i^{12}\succeq 0$. For $N_{\xi}= 2$ and $d=2$ in (\ref{eq:LMI invariance condition 3 for case 1}), the obtain same necessary conditions as in the case when $d=0$ but now for sufficiency we need to satisfy $2M_i^{11}+M_i^{12}\succeq 0,\,2M_i^{22}+M_i^{12}\succeq0, M_i^{11}+M_i^{22}+2M_i^{12}\succeq0$. 

%In the sequel, we derive sufficient LMI conditions for obtained parameter dependent matrix inequality conditions using Polya's relaxation.

\section{Tractable LMI Feasibility Conditions}\label{sec:LMI cond}
%At this point, we would direct readers to \textbf{Lemma~\ref{lemma:polya relexation}} presented in \textbf{Appendix~\ref{Appendix:polya relaxation}}, which will be extensively used in the subsequent derivations of feasibility conditions. We will need following modified multinomial coefficients, which were originally defined in \textbf{Appendix~\ref{Appendix:polya relaxation}},
%$\mathfrak{X}^{i}_{q}(d,a)=d!/(\beta_1!\cdots (\beta_i-a)!\cdots \beta_r!)$ if $\beta_i-a\in \mathbb{Z}_+$ otherwise $0$, and $\mathfrak{X}^{ij}_{q}(d,a,b)=d!/(\beta_1!\cdots (\beta_i-a)!\cdots(\beta_j-b)!\cdots \beta_r!)$ if $(\beta_i-a),(\beta_j-b)\in \mathbb{Z}_+$ otherwise 0.
The matrix inequality conditions for invariance, system constraints and performance derived in \textbf{Lemma \ref{lemma: parameter dependent invariance conditon}} and  \textbf{Lemma \ref{lem:H_2}} are nonlinear and dependent on $\xi$. Hence, solving them in the current form can be intractable.
We resolve the nonlinearity in $\bm{P}^{k,l}$ (see \eqref{eq:Pkl matrix}) by fixing the matrices $P^k=P_0^k=P_{init}$, $k=1,\ldots,N_\xi$, where $P_{init}$ is some \emph{known} matrix. As explained in \textbf{Remark~\ref{rem:linearization choice}}, we can thus allow matrices $\bar{\Lambda}_i=\Lambda_i$, $\bar{\Pi}_j=\Pi_j$, $\bar{\Upsilon}=\Upsilon$ (their ideal choices). In the following theorem, we present one of the main result of this paper which gives tractable LMI feasibility conditions for \textbf{Problem~\ref{Problem Formulation for invariance condition 1}}.
%\MMm{Perhaps for more clarity, we can state a line here,  the following theorem  is obtained by application of Polya's relaxation to polynomials in Lemma 3,4,5 }
\begin{theorem2}\label{theorem:feasibility of invariance set problem 1}
Let $\mathcal{P}(\xi)=\mathcal{P}_0(\xi)=P_{init}$ be a given matrix, then  \textbf{Problem~\ref{Problem Formulation for invariance condition 1}} has a feasible solution if,
\begin{itemize}
    \item[$i$.] there exist matrices $W$, $\bar{K}^k$, $V_i^k$, $X_i$, diagonal semi-definite matrices $\Lambda_i$, $\Gamma_i$, $\Pi_j$ and scalar $\phi_i>0$, where $k=1,\cdots,N_{\xi}$, $i=1,\ldots,n_p$ and $j=1,\ldots,n_h$  satisfying:\\
    \begin{subequations}\label{eq:invariance condition 1}
\begin{align}\label{eq:LMI invariance condition 1 for case 1}
\begin{bmatrix}
\mathsf{He}(W^{T}Y_i)-Y_i^TX_{i}Y_i & *\\
\phi_{i}e_{i}^{T}P_{init} & \phi_{i}
\end{bmatrix}&\succ 0,\\\label{eq:LMI invariance condition 2 for case 1}
\phi_i-\bm{1}^T\Lambda_i\bm{1}-\bm{1}^T\Gamma_i \bm{1}&\succ 0,
\end{align}
\begin{equation}\label{eq:LMI invariance condition 3 for case 1}
%\bm{\mathcal{M}}_i(d,q)\triangleq\sum_{k=1}^{N_{\xi}}\mathfrak{X}^{k}_{q}(d,2)\bm{M}_{i}^{k,k}(\Lambda_i,\Lambda_i)\\+\sum_{k=1}^{N_{\xi}-1}\sum_{l=k+1}^{N_{\xi}}\mathfrak{X}^{kl}_{q}(d,1,1)(\bm{M}_{i}^{k,l}(\Lambda_i,\Lambda_i)\\+\bm{M}_{i}^{l,k}(\Lambda_i,\Lambda_i)) \succ 0, \\
%\ \ q=1,\ldots, \mathfrak{L}(d+2,N_{\xi}),
\left. 
\begin{matrix}
\hspace{2.2cm}\bm{M}_{i}^{k,k}({\Lambda}_i,\Lambda_i)\succ0,&\hspace{-0.7cm}k=1,\ldots,N_{\xi}\\ 
\bm{M}_{i}^{k,l}({\Lambda}_i,\Lambda_i)\!+\!\bm{M}_{i}^{l,k}({\Lambda}_i,\Lambda_i)\succ 0, & k=1,\ldots,N_{\xi}\!-\!1,\\& l=k+1,\ldots,N_{\xi}
\end{matrix}\right\},
\end{equation}
\end{subequations}
\begin{equation}\label{eq:LMI system Constraints for case 1}
\left. 
\begin{matrix}
\hspace{2.2cm}\bm{R}_{i}^{k,k}({\Pi}_j,{\Pi}_j)\succeq0,&\hspace{-0.7cm}k=1,\ldots,N_{\xi}\\ 
\bm{R}_{i}^{k,l}({\Pi}_j,{\Pi}_j)\!+\!\bm{R}_{i}^{l,k}({\Pi}_j,{\Pi}_j)\succeq 0, & k=1,\ldots,N_{\xi}\!-\!1,\\& l=k+1,\ldots,N_{\xi}
\end{matrix}\right\},% \bm{\mathcal{R}}_j(d,q)\triangleq\sum_{k=1}^{N_{\xi}}\mathfrak{X}^{k}_{q}(d,2)\bm{R}_{j}^{k,k}(\Pi_j,\Pi_j)\!\\+\!\sum_{k=1}^{N_{\xi}-1}\sum_{l=k+1}^{N_{\xi}}\mathfrak{X}^{kl}_{q}(d,1,1)(\bm{R}_{j}^{k,l}(\Pi_j,\Pi_j)\\+\bm{R}_{j}^{l,k}(\Pi_j,\Pi_j))
% \succ 0,\\
%q=1,\ldots,\mathfrak{L}(d+2,N_\xi).
\end{equation}
to fulfill conditions \eqref{eq:invariance Condn in defn} and \eqref{eq:State Constraint in defn}.
\item[$ii$.] there exist $W$, $\bar{K}^k$, $Q^k$, $Z^k$, $F^k$ and $\Upsilon$, where $k=1,\cdots,N_{\xi}$ for a given performance bound $\gamma$ satisfying 
\begin{subequations}\label{eq:LMI H2 cost constraints case 1}
\begin{equation}\label{eq:LMI H2 cost constraint 1 case 1}
    \left. 
\begin{matrix}
\hspace{1.2cm}\bm{N}_{i}^{k,k}\succeq0,&\hspace{-0.7cm}k=1,\ldots,N_{\xi}\\ 
\bm{N}_{i}^{k,l}+\bm{N}_{i}^{l,k}\succeq 0, & k=1,\ldots,N_{\xi}-1,\\& l=k+1,\ldots,N_{\xi}
\end{matrix}\right\},
\end{equation}
\begin{equation}\label{eq:LMI H2 cost constraint 2 case 1}
\left. 
\begin{matrix}
\hspace{2.2cm}\bm{L}_{i}^{k,k}({\Upsilon},{\Upsilon})\succeq0,&\hspace{-0.7cm}k=1,\ldots,N_{\xi}\\ 
\bm{L}_{i}^{k,l}({\Upsilon},{\Upsilon})\!+\!\bm{L}_{i}^{l,k}({\Upsilon},{\Upsilon})\succeq 0, & k=1,\ldots,N_{\xi}\!-\!1,\\& l=k+1,\ldots,N_{\xi}
\end{matrix}\right\},
\end{equation}
%\hspace{4cm}$q=1,\ldots,\mathfrak{L}(d+2,N_\xi).$
\end{subequations}
to fulfill condition \eqref{eq:H2 cost}.
\end{itemize}
 An RCI set can then be obtained as in (\ref{eq:invariant set description}) and the PDCL $\mathcal{K}(\xi)=\bar{\mathcal{K}}(\xi)W^{-1}$. 
\end{theorem2}
\vspace{-0.2cm}
% Please refer to \textbf{Theorem~6} in \cite{ag21ARXIV} for the proof. 
 \begin{pf}
 \begin{itemize}
 \item[$i.$] Considering $\mathcal{P}(\xi)=\mathcal{P}_0(\xi)=P_{init}$, \eqref{eq:LMI invariance condition 1 for case 1} and \eqref{eq:LMI invariance condition 2 for case 1} are directly obtained from \eqref{eq:parameter dependent invariance condition 1} and \eqref{eq:parameter dependent invariance condition 2}, respectively. Next, we consider \eqref{eq:parameter dependent invariance condition 3}, which is a homogeneous matrix valued polynomial of degree $2$ and choose $\bar{\Lambda}_i=\Lambda_i$. 
 The \emph{l.h.s} of \eqref{eq:parameter dependent invariance condition 3} is a matrix valued polynomial in $\xi_k,\,k=1,\ldots,N_\xi$. Since $\xi_k\geq 0$, a sufficient condition for \eqref{eq:parameter dependent invariance condition 3} can be obtained by imposing each coefficient matrix of the polynomial to be positive-definite, which is given by \eqref{eq:LMI invariance condition 3 for case 1}. Similarly,  
 letting $\bar{\Pi}_j=\Pi_j$,  a sufficient condition for \eqref{eq:parameter dependent system Constraints} is \eqref{eq:LMI system Constraints for case 1}.
 \item[$ii$. ] We can prove \eqref{eq:LMI H2 cost constraint 1 case 1} and \eqref{eq:LMI H2 cost constraint 2 case 1} are sufficient for \eqref{eq:parameter dependent H2 cost constraint 1} and \eqref{eq:parameter dependent H2 cost constraint 2} by using similar arguments as mentioned in part-$i$. Notice that in \eqref{eq:LMI H2 cost constraint 2 case 1} we substitute $\bar{\Upsilon}=\Upsilon$.
 \end{itemize}%\hfill{}  \qed
 \end{pf}
\vspace{-0.2cm}
Note that, even if $P^k$'s are assumed to be constant in \textbf{Theorem~\ref{theorem:feasibility of invariance set problem 1}}, the variable matrix $W$ allows to reshape the RCI set. A similar construction to find initial RCI set was also proposed in \cite{lc19,ag19a}.
 \begin{remark2}\label{rem:zeroth order polya}
% %Since $\xi$ belongs to unit simplex $\bm{\Xi}$,. 
 The LMI conditions presented in \textbf{Theorem~\ref{theorem:feasibility of invariance set problem 1}} are sufficient conditions for invariance, system constraints and performance. From \cite{or05}, it is known that the conservatism of these LMI conditions can be reduced by applying \emph{Pólya} relaxation, which comes at a cost of increased number of total LMI conditions depending upon the order of relaxation. In fact
 the LMI conditions presented in Theorem~\ref{theorem:feasibility of invariance set problem 1} are equivalent to zeroth order Polya's relaxation. Just for the simplicity of exposition, we omit discussions related to higher order relaxation in this work. Interested readers are referred to \cite{ag21ARXIV} for higher order relaxations.
 \end{remark2} 
We  formulate feasibility conditions for \textbf{Problem~\ref{Problem Formulation for invariance condition 2}} in the next theorem. {In the theorem, matrices $P^k$'s are treated as variables and thus, inline with \textbf{Remark~\ref{remark:linearization}}, we fix $\bar{\Lambda}_i=\Lambda_i^0$, $\bar{\Pi}_j=\Pi_j^0$, $\bar{\Upsilon}=\Upsilon^0$.}

\begin{theorem2}\label{theorem:feasibility of invariance set problem 2}
Let $\mathcal{P}_0(\xi)$ and $W$ be given matrices, then  \textbf{Problem~\ref{Problem Formulation for invariance condition 2}} has a feasible solution if,
\begin{itemize}
\item[$i$.] there exist matrices $P^k$, $\bar{K}^k$, $V_i^k$, $X_i$, diagonal semi-definite matrices $\Lambda_i$, $\Gamma_i$, $\Pi_j$ and scalar $\phi_i>0$, where $k=1,\cdots,N_{\xi}$, $i=1,\ldots,n_p$ and $j=1,\ldots,n_h$  satisfying:

\begin{subequations}\label{eq:invariance condition 2}
\begin{align}\label{eq:LMI invariance condition 1 for case 2}
\begin{bmatrix}
\mathsf{He}(W^{T}Y_i)-Y_i^TX_{i}Y_i & *\\
e_{i}^{T}P^k & \phi_{i}^{-1}
\end{bmatrix}&\succ 0,\\\label{eq:LMI invariance condition 2 for case 2}
\begin{bmatrix} 
\phi_i^{-1}-\bm{1}^T\bar{\Gamma}_i \bm{1} & *\\
\phi_i^{-1}\bm{1}&\Lambda_i^{-1}
\end{bmatrix}&\succeq 0,
\end{align}\vspace{-0.25cm}
\begin{equation}\label{eq:LMI invariance condition 3 for case 2}
\left. 
\begin{matrix}
\hspace{2.2cm}\bm{\bar{M}}_{i}^{k,k}(\Lambda_i^0,\Lambda_i)\succ0,&\hspace{-0.7cm}k=1,\ldots,N_{\xi}\\ 
\bm{\bar{M}}_{i}^{k,l}(\Lambda_i^0,\Lambda_i)\!+\!\bm{\bar{M}}_{i}^{l,k}(\Lambda_i^0,\Lambda_i)\succ 0, & k=1,\ldots,N_{\xi}\!-\!1,\\& l=k+1,\ldots,N_{\xi}
\end{matrix}\right\}
\end{equation}
\end{subequations}
\begin{equation}\label{eq:LMI system Constraints for case 2}
\left. 
\begin{matrix}
\hspace{2.2cm}\bm{R}_{j}^{k,k}(\Pi_j^0,\Pi_j)\succeq0,&\hspace{-0.7cm}k=1,\ldots,N_{\xi}\\ 
\bm{R}_{j}^{k,l}(\Pi_j^0,\Pi_j)\!+\!\bm{R}_{j}^{l,k}(\Pi_j^0,\Pi_j)\succeq 0, & k=1,\ldots,N_{\xi}\!-\!1,\\& l=k+1,\ldots,N_{\xi}
\end{matrix}\right\}
 %\bm{\mathcal{R}}_j(d,q)\triangleq\sum_{k=1}^{N_{\xi}}\mathfrak{X}^{k}_{q}(d,2)\bm{R}_{j}^{k,k}(\Pi_j^0,\Pi_j)\!\\+\!\sum_{k=1}^{N_{\xi}-1}\sum_{l=k+1}^{N_{\xi}}\mathfrak{X}^{kl}_{q}(d,1,1)(\bm{R}_{j}^{k,l}(\Pi_j^0,\Pi_j)\\+\bm{R}_{j}^{l,k}(\Pi_j^0,\Pi_j))
% \succ 0,\\
%q=1,\ldots,\mathfrak{L}(d+2,N_\xi).
\end{equation}
to fulfill conditions \eqref{eq:invariance Condn in defn} and \eqref{eq:State Constraint in defn}, where  
\begin{equation}\label{eq:Mbar_kl}
\bm{\bar{M}}_{i}^{k,l}(\Lambda_i^0,\Lambda_i)\!=\!\begin{bmatrix}
\bm{P}^{k,l}(\Lambda_i^0,\Lambda_i) & * & * & *\\ 
0 &  G^T \bar{\Gamma}_i G & *&*\\
A^kW\!+\!B^k\bar{K}^l &  \phi_i^{-1}E^k &\mathsf{He}(V_i^k) & *\\
0 &  0 & V_i^k & X_i
\end{bmatrix}
\end{equation}
\item[$ii$.] there exist $P^k$, $\bar{K}^k$, $Q^k$, $Z^k$, $F^k$ and $\Upsilon$, where $k=1,\cdots,N_{\xi}$ for a given performance bound $\gamma$ satisfying 
\begin{subequations}\label{eq:LMI H2 cost constraints case 2}
\begin{equation}\label{eq:LMI H2 cost constraint 1 case 2}
\left. 
\begin{matrix}
\hspace{1.2cm}\bm{N}_{i}^{k,k}\succeq0,&\hspace{-0.7cm}k=1,\ldots,N_{\xi}\\ 
\bm{N}_{i}^{k,l}+\bm{N}_{i}^{l,k}\succeq 0, & k=1,\ldots,N_{\xi}-1,\\& l=k+1,\ldots,N_{\xi}
\end{matrix}\right\}
\end{equation}
\begin{equation}\label{eq:LMI H2 cost constraint 2 case 2}
\left. 
\begin{matrix}
\hspace{2.2cm}\bm{L}^{k,k}(\Upsilon^0,\Upsilon)\succeq0,&\hspace{-0.7cm}k=1,\ldots,N_{\xi}\\ 
\bm{L}^{k,l}(\Upsilon^0,\Upsilon)\!+\!\bm{L}^{l,k}(\Upsilon^0,\Upsilon)\succeq 0, & k=1,\ldots,N_{\xi}\!-\!1,\\& l=k+1,\ldots,N_{\xi}
\end{matrix}\right\}
%\bm{\mathcal{L}}(d,q)\triangleq\sum_{k=1}^{N_{\xi}}\mathfrak{X}^{k}_{q}(d,2)\bm{L}^{k,k}(\Upsilon^0,\Upsilon)\\+\sum_{k=1}^{N_{\xi}-1}\sum_{l=k+1}^{N_{\xi}}\mathfrak{X}^{kl}_{q}(d,1,1)(\bm{L}^{k,l}(\Upsilon^0,\Upsilon)\\+\bm{L}^{l,k}(\Upsilon^0,\Upsilon))\succ 0.
\end{equation}
\end{subequations}
to fulfill condition \eqref{eq:H2 cost}.
\end{itemize}
A PD-RCI set can then be obtained as in (\ref{eq:invariant set description}) and the PDCL is $\mathcal{K}(\xi)=\bar{\mathcal{K}}(\xi)W^{-1}$.
\end{theorem2}
%{For the proof refer to \cite{ag21ARXIV}[\textbf{Theorem~$8$}].} 
 \begin{pf}
 \begin{itemize}
 \item[$i.$] We obtain \eqref{eq:LMI invariance condition 1 for case 2} from \eqref{eq:parameter dependent invariance condition 1} by application of congruence transform and since the resultant matrix inequality is affinely dependent on the parameter. Using Schur complement  on \eqref{eq:parameter dependent invariance condition 2} and substituting $\bar{\Gamma}_i=\phi_i^{-2}\Gamma_i$, we get \eqref{eq:LMI invariance condition 2 for case 2}. By replacing $\Gamma_i=\phi_i^{2}\bar{\Gamma}_i$ and $\bar{\Lambda}_i=\Lambda_i^{0}$ in \eqref{eq:parameter dependent invariance condition 3}, %and applying congruence transform, we get
 \begin{align}\label{eq:parameter dependent matrix inequality condition for invariance 3 for case 2}
&\sum_{k=1}^{N_{\xi}} \! \xi_k^2 \bm{\bar{M}}_{i}^{k,k}(\Lambda_i^0,\Lambda_i) + \nonumber\\
&\!\!\! \sum_{k=1}^{N_{\xi}-1}\!\!\sum_{l=k+1}^{N_{\xi}} \!\!\xi_k \xi_l(\bm{\bar{M}}_{i}^{k,l}(\Lambda_i^0,\Lambda_i)\!\!+\!\!\bm{\bar{M}}_{i}^{l,k}(\Lambda_i^0,\Lambda_i)) \! \succ \! 0,
 \end{align}
  where $\bm{\bar{M}}_{i}^{k,l}$ is given in \eqref{eq:Mbar_kl}. Since \eqref{eq:parameter dependent matrix inequality condition for invariance 3 for case 2} is homogeneous matrix valued polynomial of degree $2$, we can now employ zeroth order Polya's relaxation %(see, Remark~\ref{rem:zeroth order polya}) 
  to obtain \eqref{eq:LMI invariance condition 3 for case 2}. Similarly, \eqref{eq:LMI system Constraints for case 2} is obtained from \eqref{eq:parameter dependent system Constraints}, by substituting $\bar{\Pi}_j=\Pi_j^{0}$ and using zeroth order Polya's relaxation.
 \item[$ii.$] We can prove \eqref{eq:LMI H2 cost constraint 1 case 2} and \eqref{eq:LMI H2 cost constraint 2 case 2} using similar approach as in part-$i$. Notice that in \eqref{eq:LMI H2 cost constraint 2 case 2}, we replace $\bar{\Upsilon}=\Upsilon^0$.
 \end{itemize}
% \hfill{}  \qed
 \end{pf}
%\vspace{-0.2cm}
By finding a feasible solution for \textbf{Problem~\ref{Problem Formulation for invariance condition 2}}, we obtain a PD-RCI set. However, the inequalities \eqref{eq:invariance condition 2}, \eqref{eq:LMI system Constraints for case 2} and \eqref{eq:LMI H2 cost constraints case 2} depend on the matrices $P_0^{k}$, $\Lambda_i^0$, $\Pi_j^0$ and $\Upsilon^0$, which are the initial guess of matrices $P^{k}$, $\Lambda_i$, $\Pi_j$ and $\Upsilon$, respectively. Finding an initial guess for these matrices is not straightforward; we thus obtain them by solving \textbf{Problem~\ref{Problem Formulation for invariance condition 1}}. {It is easy to verify that using solutions from \textbf{Problem~\ref{Problem Formulation for invariance condition 1}} to initialize \textbf{Problem~\ref{Problem Formulation for invariance condition 2}}, always preserves feasibility of solutions, see \textbf{Remark~\ref{remark:linearization}}.} Finally, for clarity of exposition, we summarize the main results presented so far and their interrelation with the help of a flowchart, as depicted in Fig.~\ref{fig:flowchart}.
%%% FLOW CHART %%%%%%%%%%%%

\tikzset{every picture/.style={line width=0.75pt}} %set default line width to 0.75pt
\begin{figure}
\centering
 %\begin{tikzpicture}[x=0.75pt,y=0.75pt,yscale=-0.89,xscale=0.9]
 \begin{tikzpicture}[x=0.75pt,y=0.75pt,yscale=-0.89,xscale=0.9]
%uncomment if require: \path (0,496); %set diagram left start at 0, and has height of 496
%Shape: Boxed Line [id:dp28592298600428934] 
\draw [line width=1.5]   (43,240) -- (90,240) ;
\draw [shift={(90,240)}, rotate = 539.5] [color={rgb, 255:red, 0; green, 0; blue, 0 }  ][line width=1.5]    (10.93,-3.29) .. controls (6.95,-1.4) and (3.31,-0.3) .. (0,0) .. controls (3.31,0.3) and (6.95,1.4) .. (10.93,3.29)   ;
%Straight Lines [id:da46184620315308056] 
\draw [line width=1.5]    (75,33.52) -- (115,33.52) ;
%Straight Lines [id:da7531473131557733] 
\draw [line width=1.5]    (205,322) -- (205,348) ;
\draw [shift={(205,348)}, rotate = 270] [color={rgb, 255:red, 0; green, 0; blue, 0 }  ][line width=1.5]    (14.21,-4.28) .. controls (9.04,-1.82) and (4.3,-0.39) .. (0,0) .. controls (4.3,0.39) and (9.04,1.82) .. (14.21,4.28)   ;
%Straight Lines [id:da07022475674358364] 
\draw [line width=1.5]    (205,425) -- (205,450) ;
\draw [shift={(205,450)}, rotate = 270] [color={rgb, 255:red, 0; green, 0; blue, 0 }  ][line width=1.5]    (14.21,-4.28) .. controls (9.04,-1.82) and (4.3,-0.39) .. (0,0) .. controls (4.3,0.39) and (9.04,1.82) .. (14.21,4.28)   ;
%Straight Lines [id:da5692251231491938] 
\draw [line width=1.5]    (75,33) -- (75,70) ;
\draw [shift={(75,70)}, rotate = 270] [color={rgb, 255:red, 0; green, 0; blue, 0 }  ][line width=1.5]    (14.21,-4.28) .. controls (9.04,-1.82) and (4.3,-0.39) .. (0,0) .. controls (4.3,0.39) and (9.04,1.82) .. (14.21,4.28)   ;
%Straight Lines [id:da8283892230885401] 
\draw [line width=1.5]    (331,35) -- (331,70) ;
\draw [shift={(331,70)}, rotate = 270] [color={rgb, 255:red, 0; green, 0; blue, 0 }  ][line width=1.5]    (14.21,-4.28) .. controls (9.04,-1.82) and (4.3,-0.39) .. (0,0) .. controls (4.3,0.39) and (9.04,1.82) .. (14.21,4.28)   ;
%Straight Lines [id:da9338477452864391] 
\draw [line width=1.5]    (293,35) -- (331,35) ;
%Shape: Boxed Line [id:dp501895991848254] 
\draw [line width=1.5]   (75,160) -- (142.64,160) ;
\draw [shift={(144.64,160)}, rotate = 539.5] [color={rgb, 255:red, 0; green, 0; blue, 0 }  ][line width=1.5]    (10.93,-3.29) .. controls (6.95,-1.4) and (3.31,-0.3) .. (0,0) .. controls (3.31,0.3) and (6.95,1.4) .. (10.93,3.29)   ;
%Shape: Boxed Line [id:dp37889791260323347] 
\draw [line width=1.5]   (331,160) -- (270,160) ;
\draw [shift={(270,160)}, rotate = 359.37] [color={rgb, 255:red, 0; green, 0; blue, 0 }  ][line width=0.75]    (10.93,-3.29) .. controls (6.95,-1.4) and (3.31,-0.3) .. (0,0) .. controls (3.31,0.3) and (6.95,1.4) .. (10.93,3.29)   ;
%Straight Lines [id:da36928424147876826] 
\draw [line width=1.5]    (75,134) -- (75,160) ;
%Straight Lines [id:da5593981376199875] 
\draw [line width=1.5]    (331,134) -- (331,160) ;
%Straight Lines [id:da08635720966816085] 
\draw [line width=1.5]    (205,182) -- (205,205) ;
\draw [shift={(205,205)}, rotate = 270] [color={rgb, 255:red, 0; green, 0; blue, 0 }  ][line width=1.5]    (14.21,-4.28) .. controls (9.04,-1.82) and (4.3,-0.39) .. (0,0) .. controls (4.3,0.39) and (9.04,1.82) .. (14.21,4.28)   ;
%Straight Lines [id:da37729787534395265] 
\draw [line width=1.5]    (205,282) -- (205,306) ;
\draw [shift={(205,306)}, rotate = 270] [color={rgb, 255:red, 0; green, 0; blue, 0 }  ][line width=1.5]    (14.21,-4.28) .. controls (9.04,-1.82) and (4.3,-0.39) .. (0,0) .. controls (4.3,0.39) and (9.04,1.82) .. (14.21,4.28)   ;
%Shape: Rectangle [id:dp7237766756520665] 
\draw  (73,195) -- (328,195) -- (328,290) -- (73,290) -- cycle ;

\draw (73,195) -- (160,195) -- (160,212) -- (73,212) -- cycle ;
%[dash pattern={on 0.84pt off 2.51pt}]
%Shape: Rectangle [id:dp3832777420792286] 
\draw (73,337) -- (328,337) -- (328,432) -- (73,432) -- cycle ;

\draw (73,337) -- (160,337) -- (160,354) -- (73,354) -- cycle ;

%Shape: Rectangle [id:dp9265821842333417] 
\draw   (115,5) -- (293,5) -- (293,72) -- (115,72) -- cycle ;

% Text Node
\draw (90,230) node [anchor=north west][inner sep=0.75pt]   [align=center] {{\small LMIs for invariance condition: \eqref{eq:invariance condition 1}}\\{\small LMIs for system constraints: \eqref{eq:LMI system Constraints for case 1}}\\{\small LMIs for Performance constraints: \eqref{eq:LMI H2 cost constraints case 1}}};
% Text Node
\draw (115,6) node [anchor=north west][inner sep=0.75pt]  [rotate=-0.13] [align=center] {\textbf{Problem formulation}\\{\small Invariance condition~\eqref{eq:invariance Condn in defn}}\\{\small System constraints~\eqref{eq:State Constraint in defn}}\\{\small Performance constraints~\eqref{eq:H2 cost} }};
% Text Node
\draw (275,100) node [anchor=north west][inner sep=0.75pt]   [align=center] {{\small Performance: \eqref{eq:parameter dependent H2 cost constraint}}};
% Text Node
\draw (42,80) node [anchor=north west][inner sep=0.75pt]   [align=center] {\textbf{Lemma~\ref{lemma: parameter dependent invariance conditon}}};
% Text Node
\draw (297,80) node [anchor=north west][inner sep=0.75pt]   [align=center] {\textbf{Lemma ~\ref{lem:H_2}}};
% Text Node
\draw (137,145) node [anchor=north west][inner sep=0.75pt]   [align=center] {\textbf{ \ }{Polya's relaxation}\\ \ \ \ \ (zeroth order)};
% Text Node
\draw (32,95) node [anchor=north west][inner sep=0.75pt]   [align=center] {{\small Invariance:  \eqref{eq:parameter dependent invariance condition}}\\{\small Constraints:  \eqref{eq:parameter dependent system Constraints}}};
% Text Node
\draw (167,213) node [anchor=north west][inner sep=0.75pt]   [align=left] {\textbf{Theorem \ref{theorem:feasibility of invariance set problem 1}}};
% Text Node
\draw (103,307) node [anchor=north west][inner sep=0.75pt]   [align=left] {\textbf{feasible solution \ \ } };
% Text Node
\draw (166,360) node [anchor=north west][inner sep=0.75pt]   [align=center] {\textbf{Theorem \ref{theorem:feasibility of invariance set problem 2}}};
% Text Node
\draw (90,375) node [anchor=north west][inner sep=0.75pt]   [align=center] {{\small LMIs for invariance condition: \eqref{eq:invariance condition 2}}\\{\small LMIs for system constraints: \eqref{eq:LMI system Constraints for case 2}}\\{\small LMIs for performance constraints: \eqref{eq:LMI H2 cost constraints case 2}}};
% Text Node
\draw (222,450) node [anchor=north west][inner sep=0.75pt]    {$:(\mathcal{P}(\xi),W,\mathcal{K}(\xi))$};
% Text Node 
\draw (97,450) node [anchor=north west][inner sep=0.75pt]   [align=left] {\textbf{feasible solution \ \ } };
% Text Node
\draw (226,307) node [anchor=north west][inner sep=0.75pt]    {$:(W,\mathcal{K}(\xi))$};
% Text Node
\draw (41,223) node [anchor=north west][inner sep=0.75pt]   [align=left] {$P_{init}$};
% Text Node
\draw (78,198) node [anchor=north west][inner sep=0.75pt]   [align=left] {\textbf{Problem \ref{Problem Formulation for invariance condition 1}}};
% Text Node
\draw (78,340) node [anchor=north west][inner sep=0.75pt]   [align=left] {\textbf{Problem \ref{Problem Formulation for invariance condition 2}}};
\end{tikzpicture}
\caption{Flowchart summarizing the main results.}
\label{fig:flowchart}
\end{figure}
\section{Iterative PD-RCI Set Computation}\label{section:algorithms for rci set computation}
Our primary goal is to compute PD-RCI set \eqref{eq:invariant set description} of desirabily large volume and the PDCL controller \eqref{eq:gain schedule feedback}. Thus, we need to formulate a method which computes a maximum volume set feasible to conditions proposed in \textbf{Theorem~\ref{theorem:feasibility of invariance set problem 1}} and  \textbf{Theorem~\ref{theorem:feasibility of invariance set problem 2}}.  {In the original form, the conditions in these theorems were nonlinear, and to make them tractable for solving, we linearized them by using \textbf{Lemma~\ref{lemma:sucessive linearization lemma}}. As mentioned in the \textbf{Remark~\ref{remark:linearization}}}, the linearization introduces conservatism, which can be reduced by adopting an iterative scheme, where   we first consider \textbf{Problem~\ref{Problem Formulation for invariance condition 1}} in which we assumed $P(\xi)=P^0(\xi)=P_{init}$. As shown in \cite{ag20}, the volume of the considered RCI set is   proportional to $|\mathsf{det}(W)|$. We next propose an optimization problem that computes desirably large RCI set for \textbf{Problem~\ref{Problem Formulation for invariance condition 1}}.

\subsection{Initial RCI set computation}\label{sec:initial rci set}
\vspace{-0.2cm}
We develop an iterative scheme in which we solve a determinant maximization problem under LMI conditions presented in \textbf{Theorem~\ref{theorem:feasibility of invariance set problem 1}}. Similar to \cite{ag19b}, we will try to iteratively maximize the volume to avoid enforcing symmetry on $W$. The basic idea is to maximize the determinant of a different matrix $J$, which is required to satisfy
\begin{equation}\label{T Condition}
W^TW\succcurlyeq J \succ 0,
\end{equation}
Condition \eqref{T Condition} ensures that $\mathsf{det}(J)\leq |\mathsf{det}(W)|^2$.
Since (\ref{T Condition}) is not an LMI, it needs to be replaced with a sufficient condition. This is done within the iterative scheme in which the solution of $W$ at the previous step is represented as $W^0$. A sufficient condition for (\ref{T Condition}) is formulated in terms of $W^0$ as (see \cite{ag19b}) 
\begin{equation}\label{eq:condition for Det Increase}
W^{T}W^0+(W^0)^{T}W-(W^{0})^{T}W^{0}\succcurlyeq J \succ 0.
\end{equation}
Note that this condition is necessarily satisfied with $W=W^0$. Thus,
maximization of $\mathsf{det}(J)$ under (\ref{eq:condition for Det Increase}) would lead to a solution $W$ that satisfies $|\mathsf{det}(W)|\geq|\mathsf{det}(W^0)|$.
Moreover, as described in \textbf{Remark~\ref{remark:linearization}}, at each iteration we update $Y_i=(X_i^0)^{-1}W^0$ in \eqref{eq:LMI invariance condition 1 for case 1}, where $X_i^0$ is previous solution of $X_i$. This allows us to develop the following iterative algorithm to compute RCI sets of increased volume at each step for a priori chosen matrix $P_{init}$
\begin{equation}\label{eq:optimization problem for initial RCI set}
\left.\begin{matrix}
\begin{array}{cc}
\max & \mathsf{log}\,\mathsf{det}(J) \\
\phi_{i},W,\bar{K}^k,V_i^k,X_i,\Lambda_{i},\Gamma_{i}  & \\
\Pi_j, Q^k, Z^k,F^k,\Upsilon, J &\\
\text{subject to:} & \hspace{-0.8cm}
\eqref{eq:invariance condition 1},\eqref{eq:LMI system Constraints for case 1},
\eqref{eq:LMI H2 cost constraints case 1}\;
\text{and}\; (\ref{eq:condition for Det Increase}) 
\end{array}
\end{matrix}\right\}
\end{equation}
\textbf{Initial Optimization to Compute $W^0$:} Condition (\ref{eq:condition for Det Increase}) is removed and  $\mathsf{log}\,\mathsf{det}(J)$ is changed to $\mathsf{log}\,\mathsf{det}(W+W^T)$;
%or $\mathsf{log}\,\mathsf{det}(W+W^T)$  (to avoid conservatism due to symmetric $W$); 
(\ref{eq:LMI invariance condition 1 for case 1}) is imposed with $Y_{i}\rightarrow I$. 
\subsection{Computation of PD-RCI sets}
%\MMm{To compute a desirably large set for \textbf{Problem~\ref{Problem Formulation for invariance condition 2}}, we need to formulate a new cost function since matrices $P^k$'s are variables, this also implies we can now fix the matrix $W$ obtained by solving \eqref{eq:optimization problem for initial RCI set}.} 
In order to compute a desirably large PD-RCI set, conforming \textbf{Problem~\ref{Problem Formulation for invariance condition 2}}, we formulate a new optimization problem. %for volume maximization of PD-RCI set treating matrices $P^k$'s as its optimization variables. 
In this problem, we fix the matrix $W$ obtained by solving \eqref{eq:optimization problem for initial RCI set}, and now treat matrices $P^k$'s as optimization variables. %Moreover, the PD-RCI set description \eqref{eq:invariant set description}  is now parameter dependent. 
By construction, for each $\xi\in \bm{\Xi}$, $\bm{\mathcal{S}}(\xi)$ is an 0-symmetric polytope in the state-space. Thus, an intuitive way to maximize the volume of such a set is to compute matrices $P^k$'s such that the sum of the volumes of each slice of $\bm{\mathcal{S}}(\xi)$ is maximized. However, maximizing infinite slices of the PD-RCI set would lead to solving semi-infinite problem, which is intractable. Nevertheless, to deal with such intractability, we only maximize the slices $\bm{\mathcal{S}}(\xi^m) = \left \{  x\in\mathbb{R}^{n_x} : -\bm{1}\!\leq\! \mathcal{P}(\xi^m)W^{-1}x \!\leq\! \bm{1} \right \}$, corresponding to the finite set of grid points $\xi^m\in\bm{\Xi},\,m=1,\ldots,N_m$. For example, a possible choice of the grid points  can be the vertices of $\bm{\Xi}$. We propose a novel volume maximization approach for polytopic sets  which leads to the following SDP problem. 
% which corresponds to $\xi=\xi^m, \ m=1,\ldots, N_{\xi}$, where $\xi^m = [\underset{m-1}{\underbrace{0,\ldots,0}}, \ 1, \ \underset{N_{\xi}-m}{\underbrace{0,\ldots,0}}]^{T}$ are the vertices of the simplex set $\bm{\Xi}$, is sufficient. 
%One indirect approach toward volume maximization could be to inscribe an ellipsoid inside the set $\bm{\mathcal{S}}$ and then maximize the volume of the ellipsoid, as done in \cite{ag19a,lc19}. This approach is known to be conservative, which can also be understood from the fact that different polytopic sets can have the same largest inscribed ellipsoid.  %Moreover, in our case, the set description is parameter dependent. Based on the result proposed in \textbf{Appendix~\ref{sec:volume maximization}}, we next present an approach which directly optimizes matrices $P^k$'s such that the resulting set $\bm{\mathcal{S}}$ has a large volume.
%Since the slices $\bm{\mathcal{S}}(\xi^m)$ are polytopes, we can utilize the volume maximization approach presented in \textbf{Appendix~\ref{sec:volume maximization}}. %The main idea is to approximate the volume of a polytope by using Monte-Carlo integration techniques and using its convex approximations.

%\MM{ADD A REMARK FOR GENERAL SLICES CASE. Write proposition only for the vertices }
\begin{prop}\label{prop:vol_max}
  Given $N_m$ number of grid points, the slices $\bm{\mathcal{S}}(\xi^m)$, $m=1,\ldots,N_m$ of desirably large volume characterizing a PD-RCI set,  can be obtained by solving the following  SDP problem in an iterative manner, 
\begin{equation}\label{eq:optimization problem for PDRCI set}
\left.\begin{matrix}
\begin{array}{cc}
\min & \sum_{n=1}^{N_\sigma}\sum_{m=1}^{N_m}  \sigma^m_n \\
\phi_{i},P^k,\bar{K}^k,V_i^k,X_i,\Lambda_{i},\Gamma_{i}  & \\
\Pi_j, Q_1^k, Z^k,F^k,\Upsilon, \sigma^m_n &\\
\text{subject to:} & \sigma^m_n\geq0,\\
&\hspace{-0.8cm}\begin{bmatrix}
\;\;\,\widetilde{P}W^{-1}\\ 
-\widetilde{P}W^{-1}
\end{bmatrix}\widetilde{x}_n-\begin{bmatrix}
\bm{1}\\ 
\bm{1}
\end{bmatrix}\leq \begin{bmatrix}
\widetilde{\sigma}\\ 
\widetilde{\sigma}
\end{bmatrix},\\
&\hspace{-0.8cm}\eqref{eq:invariance condition 2},\eqref{eq:LMI system Constraints for case 2}\;\text{and}\;
\eqref{eq:LMI H2 cost constraints case 2}. 
\end{array}
\end{matrix}\right\}
\end{equation}
where $\widetilde{P}=[\mathcal{P}(\xi^1)^T,\cdots, (\mathcal{P}(\xi^{N_m})^T]^T \in \mathbb{R}^{(n_p N_{m}) \times n_x}$,  $\widetilde{\sigma}=[\sigma^{1}_{n}\mathbf{1},\cdots, \sigma^{N_m}_{n}\mathbf{1}]^T \in \mathbb{R}^{n_p N_{m}}$ and $\{\widetilde{x}_n\}_{n=1}^{N_\sigma}$  are the vertices of 
  some known $n_x$ dimensional outer bounding box $\bm{\mathfrak{B}}$ which contains the state  constraint set $\bm{\mathcal{X}}$.
\end{prop}
\begin{pf}
 We refer the reader to \textbf{Appendix~\ref{sec:volume maximization}} for the details of the volume maximization algorithm which is based on Monte-Carlo technique presented in \cite{ben2016, PigaSetmem19}. The convex cost function formulated based on the algorithm presented in \textbf{Appendix~\ref{sec:volume maximization}}, is combined with LMI conditions \eqref{eq:invariance condition 2}, \eqref{eq:LMI system Constraints for case 2} and \eqref{eq:LMI H2 cost constraints case 2} for invariance, system  and performance constraints respectively, giving an SDP problem \eqref{eq:optimization problem for PDRCI set}. \hfill{}  \qed
 \end{pf}

Assuming that the initial values of $\mathcal{P}_0,  W,Y_i,\Lambda_i^0,\Gamma_j^0$ and $\Upsilon^0$ are available after solving \eqref{eq:optimization problem for initial RCI set}, we summarize the whole approach to compute PD-RCI set in \textbf{Algorithm~\ref{alg:pdrci set}}.  As a consequence of the adopted successive linearization approach (see \textbf{Remark~\ref{remark:linearization}}), \textbf{Algorithm~\ref{alg:pdrci set}} always has a feasible solution at the first iteration if initialized using solutions from \eqref{eq:optimization problem for initial RCI set}.
%We have already shown that \textbf{Algorithm~\ref{alg:pdrci set}} always has a feasible solution at the first iteration if initialized using solutions from \eqref{eq:optimization problem for initial RCI set}. 
{The update scheme in the algorithm alleviates the conservatism introduced while linearizing the equation \eqref{eq:LMI invariance condition 1 for case 2}, \eqref{eq:LMI invariance condition 3 for case 2}, \eqref{eq:LMI system Constraints for case 2} and \eqref{eq:LMI H2 cost constraint 2 case 2} using \textbf{Lemma~\ref{lemma:sucessive linearization lemma}}.} The systematic update procedure also guarantees that the solutions from the previous iteration are feasible in the current iteration. Thus, at each iteration we find a new PD-RCI set of larger volume until the specified number of iterations are performed, or convergence is achieved. We purposely present the termination of the algorithm based on the number of iteration instead of convergence to emphasize that latter is not necessary.

\begin{algorithm}[t!]
  \caption{:~Computing PD-RCI set.}
  \label{alg:pdrci set}
  \begin{algorithmic}
    \State \textbf{Input:} System \eqref{eq:lpv system}, $\bm{\mathcal{X}_u}$, $\bm{\mathcal{W}}$, $\mathcal{P}_0$, $W$, $Y_i$ $\Lambda_i^0$, $\Gamma_j^0$, $\Upsilon^0$
    \State \textbf{Output:} $\mathcal{P}(\xi)$, $\mathcal{K}(\xi)$
    \While{$\mathrm{Iteration}\geq$ 0}
    \vspace{0.2cm}
    \State [$\mathcal{P}$,$\bar{\mathcal{K}}$,$X_i$,$\Lambda_i$, $\Gamma_j$, $\Upsilon$] $\gets \mathrm{solve}~\eqref{eq:optimization problem for PDRCI set}$
    \vspace{0.2cm}
    \State \textbf{\textit{Update:}} $Y_i\gets X_i^{-1}W$,  $\mathcal{P}_0\gets\mathcal{P}$, $\Lambda_i^0\gets \Lambda_i$,\\\hspace{1.9cm}  $\Gamma_j^0\gets \Gamma$, $\Upsilon^0 \gets\Upsilon$\\\vspace{0.2cm}\hspace{0.5cm}$\mathrm{Iteration}\gets\mathrm{Iteration}-1$
    \vspace{0.2cm}\EndWhile
  \end{algorithmic}
\end{algorithm}

%  {\color{red} 
\begin{remark2}
 We remark that in \textbf{Proposition~\ref{prop:vol_max}}, we maximize the volume of a finite set of slices corresponding to a chosen set of grid points in the scheduling parameter space. Here, a trade-off is expected to be achieved, i.e., the finer the grid,  more likely the initial measured value of the scheduling will correspond to the one selected in the grid points, in order to obtain a larger volume set, at the cost of higher computational complexity. %If the initial scheduling parameter $\xi(0)$ does not belong to the chosen set of grid points, then the corresponding set is not optimized for volume, although still guaranteeing the invariance.
 Alternatively, in order to reduce the computational complexity, one may even choose to only maximize the volume of the intersection of the slices corresponding to the vertices of the scheduling parameter.  
   \end{remark2}

Next, we address some known implementation issues and some workarounds which could potentially help to overcome it.

 \subsection{Practical Issues}
 \subsubsection{Computational complexity}
 As explained in the previous section, we compute PD-RCI set by sequentially solving SDP \eqref{eq:optimization problem for initial RCI set} to obtain initial solution and \eqref{eq:optimization problem for PDRCI set} in \textbf{Algorithm~\ref{alg:pdrci set}}. 
 The computation complexity of both the SDPs are approximately similar. The SDP \eqref{eq:optimization problem for PDRCI set} consist of $n_p\times\left(N_{\xi}+1+\frac{(N_{\xi}+1)N_{\xi}}{2}\right)$ LMI constraints for invariance \eqref{eq:invariance condition 2}, $n_h\times\frac{(N_{\xi}+1)N_{\xi}}{2}$ LMI constraints for system constraints \eqref{eq:LMI system Constraints for case 2}, and $2\times\frac{(N_{\xi}+1)N_{\xi}}{2}$ LMI constraints for performance \eqref{eq:LMI H2 cost constraints case 2}. Furthermore, for volume maximization, we introduced $2n_p N_\xi\times N_\sigma$ affine inequality constraints. All the constraints are in terms of $(3n_p+N_{\xi}(8+n_p)+n_h)$ matrix variables and $(n_p+N_{\sigma}\times N_m)$ scalar variables. The computational complexity can be largely impacted by choice of representational complexity $n_p$. %and the number of sample points $N_\sigma$. 
 Theoretically, $n_p$ should be as large as possible to have the least conservative formulation. By choosing $n_p$, a trade-off can be achieved between computational complexity and conservatism. %One strategy could be to keep $d$ small initially and select matrix $P_{init}$ and thus $n_p$ that represents a full dimensional bounded polytope.% Moreover, as pointed out in \textbf{Remark}~\ref{rem:choice_of_sample}, $N_\sigma$ can be made small by selecting only the vertices of the box $\bm{\mathfrak{B}}$ as samples.
 Another strategy to reduce computational complexity, again at the cost of conservatism, could be to reduce the number of system vertices $N_\xi$. We can reduce $N_\xi$ by modifying the system description such that the trajectories of the modified system over-bounds the trajectories of the original. However, it may not always be possible to reduce $N_\xi$. 
\subsection{Computation of the RCI set for quasi-LPV systems}\label{sec:quasi LPV}
If the scheduling parameters $\xi$ are function of system states $x$ and input $u$, then the system is called as \emph{quasi-LPV} (qLPV).
%In the case of qLPV systems, the scheduling parameters $\xi(t)$ are the function of the system states $x(t)$, and
%For qLPV systems, initial $\xi(t)$ cannot be selected independently from the $x(t)$ and $u(t)$. 
For qLPV systems, we need to keep the RCI set description independent of parameter i.e., by restricting $P^k=P,\;\forall k=1,\ldots,N_\xi$. Alternatively, we can construct a set $\bm{\Breve{\mathcal{S}}} = \bigcap_{\forall\xi\in\bm{\Xi}} \bm{\mathcal{S}}(\xi)$. Notice that the set $\bm{\Breve{\mathcal{S}}}$ also satisfies conditions \eqref{eq:invariance Condn in defn} and \eqref{eq:State Constraint in defn} if $\bm{\mathcal{S}}(\xi)$ satisfies them, and since it is independent of $\xi$, we call it simply RCI set. The set $\bm{\Breve{\mathcal{S}}}$ can be possibly larger (volume-wise) than the set obtained by restricting $P^k=P$, due to extra degree of freedom provided by additional variables involved in the overall optimization problem when computing the former. Even though we define $\bm{\Breve{\mathcal{S}}}$ as the intersection of infinite slices of $\bm{\mathcal{S}}(\xi)$, 
% we have proved that it can be exactly obtained by intersecting the slices generated at vertices $\xi^m, m = 1, \ldots , N_\xi$ of the set $\bm{\Xi}$ (see, \cite{ag21ARXIV}), \emph{i.e.},\vspace{-0.2cm}
% \begin{equation}\label{eq:RCI set with finite intersection}
%     \bm{\Breve{\mathcal{S}}} \triangleq \bigcap_{\forall\xi \in\bm{\Xi}} \bm{\mathcal{S}}(\xi)= \bigcap_{\forall\xi^m\in\bm{\Xi}} \bm{\mathcal{S}}(\xi^m)\vspace{-0.5cm}
% \end{equation}
 we will next prove that it can be accurately obtained by performing intersections of only a few selected slices. More specifically, we will prove that\vspace{-0.2cm}
 \begin{equation}\label{eq:RCI set with finite intersection}
     \bm{\Breve{\mathcal{S}}} \triangleq \bigcap_{\forall\xi \in\bm{\Xi}} \bm{\mathcal{S}}(\xi)= \bigcap_{\forall\xi^m\in\bm{\Xi}} \bm{\mathcal{S}}(\xi^m),
 \end{equation}
where $\xi^m,\;m=1,\ldots,N_\xi$ are the vertices of the set $\bm{\Xi}$. Since $\xi^m\in \bm{\Xi}$, to prove \eqref{eq:RCI set with finite intersection}, it is enough to prove 
 \begin{equation}\label{eq:finite intersection}
 \bigcap_{\forall\xi^m\in\bm{\Xi}} \bm{\mathcal{S}}(\xi^m) \subseteq \bigcap_{\forall\xi \in\bm{\Xi}} \bm{\mathcal{S}}(\xi).    
 \end{equation}
% Given a set of form \eqref{eq:invariant set description}, we can show that the relation \eqref{eq:finite intersection} is always true, see \cite{ag21ARXIV} for the proof.}

 \begin{lemma2}\label{lem:finite intersection}
 Given a set of form \eqref{eq:invariant set description}, the relation \eqref{eq:finite intersection} is always true.
 \end{lemma2}
 \begin{pf}
 Consider any point $x\in \bigcap_{\forall\xi^m\in\bm{\Xi}} \bm{\mathcal{S}}(\xi^m)$, to complete the proof it is sufficient to show that $x\in \bigcap_{\forall\xi \in\bm{\Xi}} \bm{\mathcal{S}}(\xi)$. Towards this, since $x\in \bigcap_{\forall\xi^m\in\bm{\Xi}} \bm{\mathcal{S}}(\xi^m)$ implies following inequalities holds element-wise
 \begin{equation}
     \underbrace{|P^mW^{-1}x|}_{y^m}\leq \bm{1},\; \forall m=1,\ldots,N_\xi.
 \end{equation}
 Thus, for any  $\xi = [\xi_1, \xi_2,\cdots,\xi_{N_{\xi}}]^T \in \bm{\Xi}$, we have
 \begin{equation}
     \left\{\sum_{m=1}^{N_\xi}\xi_m y^m \!\leq\! \bm{1}\right\} \Rightarrow 
     \left\{\left |\sum_{m=1}^{N_\xi} \xi_m P^m W^{-1}x\right |\!\leq\! \bm{1}\right\}.
 \end{equation}
Since above relation holds $\forall \xi \in \bm{\Xi}$, hence \eqref{eq:finite intersection} holds.\hfill{}  \qed
 \end{pf}
 
\section{Numerical Examples}\label{sec:examples}
We now demonstrate the potential of the proposed algorithm through examples.
The algorithm is implemented in Matlab on a Intel Core i7-555U CPU with 8 GB RAM using YALMIP \citep{yalmip} and the solver SeDuMi.%~\citep{sedumi}. %The volumes of polytopes are computed  using MPT \cite{MPT3}.
\vspace{-0.2cm}
\subsection{Double Integrator}
\begin{figure}
    \centering
    \includegraphics[scale=0.52]{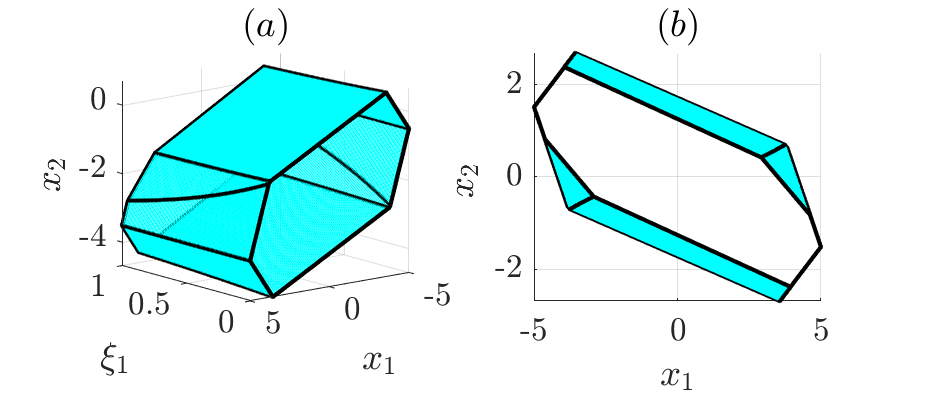}
    \caption{$(a)$ Plot of the PD-RCI set $\bm{\mathcal{S}}(\xi)$ in \eqref{eq:invariant set description} w.r.t $\xi_1$ and $(b)$ projection $\bm{\mathcal{S}}(\xi)$ on $(x_1,x_2)$ axis.}
    \label{fig:3d plot of double integrator}
\end{figure}

\begin{figure}
    \centering
    \begin{subfigure}{0.2\textwidth}
      \hspace{-1.2cm}
    \includegraphics[scale=0.23]{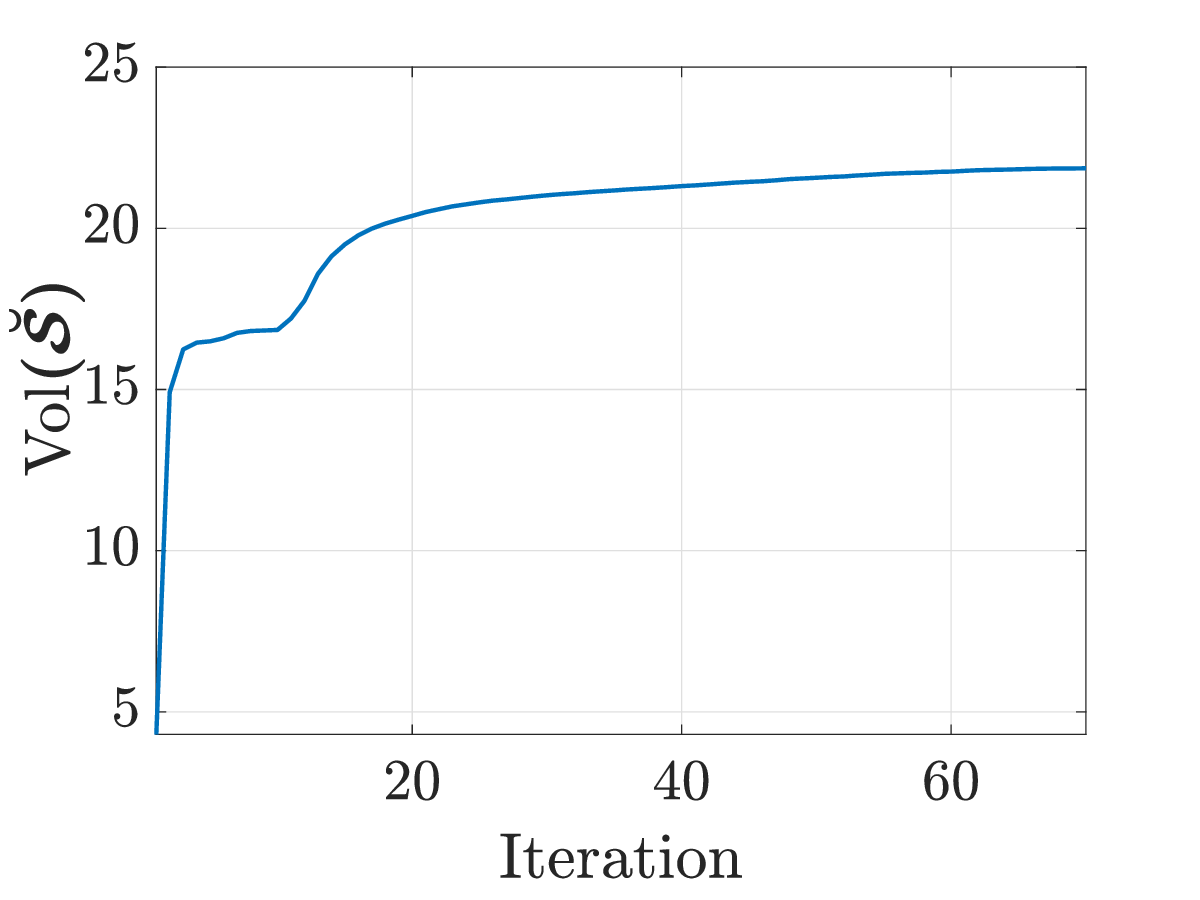}
        \caption{}\label{fig:volume double integrator}
    \end{subfigure}  
      %\hspace{0.8cm}
  \begin{subfigure}{0.2\textwidth}
 \includegraphics[scale = 0.39]{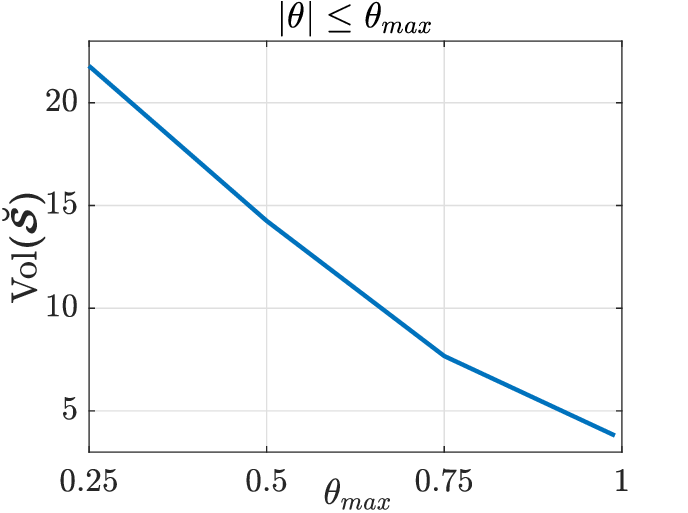}
    \caption{} \label{fig:parameter_var}
\end{subfigure}  
    \caption{{Volume of the set $\bm{\Breve{\mathcal{S}}}$ plotted against $(a)$ iteration, $(b)$   different scheduling parameter bounds.} }  \end{figure}

\begin{figure}
    \centering
    \includegraphics[width = 5.2cm, height = 4cm]{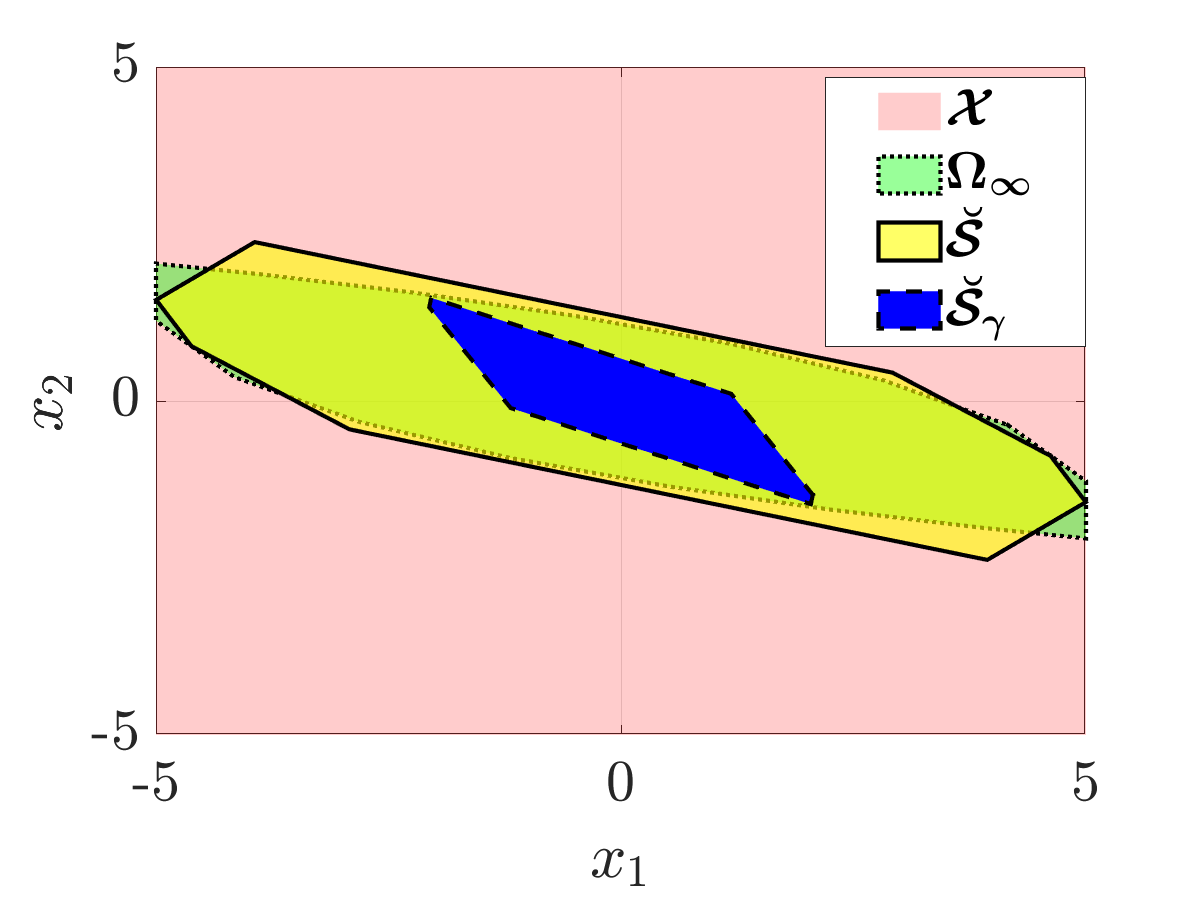}
    \caption{Admissible set $\bm{\mathcal{X}}$ (red), maximal RCI set $\bm{\Omega}_{\infty}$ (green), and RCI sets $\bm{\Breve{\mathcal{S}}}$ (yellow) and $\bm{\Breve{\mathcal{S}}}_{\gamma}$ (blue).}
    \label{fig:maximal invariant set for double integrator}
\end{figure}

%For a better visualisation of the PD-RCI set, 
Let us consider a parameter-varying double integrator: \vspace{-0.2cm}
\begin{equation}\label{eq:double integrator system}
x^+\!=\!\begin{bmatrix}
1\!+\!\theta & 1\!+\!\theta\\ 
0 & 1\!+\!\theta
\end{bmatrix}x\!+\!\begin{bmatrix}
0\\ 
1\!+\!\theta
\end{bmatrix}u\!+\!\begin{bmatrix}
1\\ 
0
\end{bmatrix}w,\vspace{-0.2cm}
\end{equation}
where $|\theta|\!\leq\! 0.25$ is a time-varying parameter. The state and control input constraints, and the disturbance bounds are expressed as $|x|\!\leq \!\left [5,\,\,5  \right ]^{T}$, $|u|\leq 1$, $|w|\leq 0.25$.
%Even though in our formulation we assume that the current values of parameters are observable, it is difficult to utilize this information in current approaches for computing RCI sets. Commonly these parameters are treated as unknown elements in the system dynamics for which RCI set are computed \cite{bp05_2,mc03}, which makes them conservative. 
We rewrite \eqref{eq:double integrator system} in the form \eqref{eq:affine lpv representation} with $N_\xi = 2$ and \vspace{-0.2cm}
\begin{equation}
\begin{bmatrix}
\begin{array}{c|c|c} 
A^1&B^1&E^1\\\hline
A^2&B^2&E^2
\end{array}
\end{bmatrix}=
%\begin{smallmatrix}
%\begin{array}{cc|c|c}
\smallmat{
\begin{array}{cc|c|c}
1.25 & 1.25 & 0 &1\\ 
   0 & 1.25 & 1.25&0\\\hline
0.75 & 0.75 & 0 &1\\ 
   0 & 0.75 & 0.75&0
 \end{array}
   },\vspace{-0.2cm}
\end{equation}
%\begin{align}
%A^1& = \begin{bmatrix}
%0.75 & 0.75\\ 
%0 &  0.75
%\end{bmatrix}, A^2 =  %\begin{bmatrix}
%1.25 & 1.25\\ 
%0 &  1.25
%\end{bmatrix},\\
%B^1& = \begin{bmatrix}
%0\\ 
%0.75
%\end{bmatrix},  B^2 = \begin{bmatrix}
%0\\ 
%1.25
%\end{bmatrix}, E^{1}=E^{2}=\mathcal{E},
%\end{align}
{where $\xi_1 \!=\! (0.25 + \theta)/0.5$ and $\xi_2 \!=\! (0.25-\theta)/0.5$. We then select $P_{init}$ as described in \cite[\textbf{Remark~1}]{ag20}
%\textbf{Remark~\ref{rem:initial choice of P}}}, 
}
and solve \eqref{eq:optimization problem for initial RCI set} iteratively until convergence, which took 10 iterations and thus obtain all the matrices needed to initialize \textbf{Algorithm~\ref{alg:pdrci set}}. Finally, the PD-RCI set $\bm{\mathcal{S}}(\xi)$, shown in Fig.~\ref{fig:3d plot of double integrator}, is obtained after performing $60$ iterations of \textbf{Algorithm~\ref{alg:pdrci set}}. The average  computation time is  $9.31$ seconds per iteration. The obtained matrices characterizing PD-RCI set and PDCL are\vspace{-0.2cm}
\begin{align*}%\nonumber
[\begin{array}{c|c}
P^1 & P^2
\end{array}]&=\begin{bmatrix}
\begin{array}{cc|cc}
   -0.4111 &  -0.1354  &  -0.3257  & -0.0854\\
\;\,0.0303 &  -0.5151  & \;\, 0.0404  & -0.3823\\
\;\,0.4867 &  -0.2474  & \;\, 0.4867  & -0.2474\\
\;\,0.4884 &  -0.0504  & \;\, 0.4883  & -0.0506
\end{array}
\end{bmatrix},\\
\begin{bmatrix}\!
\begin{array}{c|c}
W&\!\begin{array}{c} K^1\\ \hline K^2\end{array} \\ 
\end{array}\!\!\end{bmatrix}&=
\begin{bmatrix}\!\!
\begin{array}{c|c}
\begin{array}{cc}
\;\,2.4373 &   -0.6691\\
   -0.7327 &\;\,0.8379
\end{array}&\! \begin{array}{cc} -0.2246 &  -0.7898\\\hline-0.1506 &  -0.5601\end{array}
\end{array} \!\!
\end{bmatrix}.\vspace{-0.2cm}
%\begin{bmatrix}
%W\\ \hline
%K^1\\ \hline
%K^2
%\end{bmatrix}&=\begin{bmatrix}
%\;\,2.4373 &  -0.6691\\
%   -0.7327  &  \;\,0.8379\\\hline
%   -0.2246 &  -0.7898\\\hline
%   -0.1506 &  -0.5601
%\end{bmatrix}.
\end{align*}
%In Fig.~\ref{fig:3d plot of double integrator} $(a)$, we plot the PD-RCI set for $\xi_1\in[0,1]$, and in Fig.~\ref{fig:3d plot of double integrator} $(b)$ we show the projection of the set on the state-space axis $(x_1,x_2)$. 
The RCI set $\bm{\Breve{\mathcal{S}}}$ in \eqref{eq:RCI set with finite intersection} can be seen in the Fig.~\ref{fig:3d plot of double integrator} $(b)$ as bounded \emph{colourless} region. The region outside the set $\bm{\Breve{\mathcal{S}}}$, highlighted in cyan, consists of points which can be brought within the RCI set $\bm{\Breve{\mathcal{S}}}$ in one step for some selectable initial value of the parameter $\theta$, thus, enlarging the overall set of safe initial states.%, which was also highlighted in \textbf{Remark~\ref{rem:classical vs pdrci set}}.
To compare the volume gain between \textbf{Problem~\ref{Problem Formulation for invariance condition 1}} and \textbf{Problem~\ref{Problem Formulation for invariance condition 2}}, we plot the volume of the set $\bm{\Breve{\mathcal{S}}}$ at each iteration, as shown in Fig.~\ref{fig:volume double integrator}. In the figure, it can be seen that there is an additional (approximately) 29\% gain in the volume when the proposed Monte-Carlo based volume maximization approach is utilized. % after the first $10$ iterations for which \textbf{Problem~\ref{Problem Formulation for invariance condition 1}} is solved.% when \textbf{Problem~\ref{Problem Formulation for invariance condition 2}} is solved after \textbf{Problem~\ref{Problem Formulation for invariance condition 1}}. 
%\begin{figure}
%    \centering
%    \includegraphics[scale=0.35]{double_integrator_volume.eps}
%    \caption{Volume of the set $\bm{\Breve{\mathcal{S}}}$ plotted %against the iteration.}
 %   \label{fig:volume double integrator}
%\end{figure}
For comparison, we plot the computed set $\bm{\Breve{\mathcal{S}}}$ and the maximal RCI set $\bm{\Omega}_{\infty}$ obtained using classical geometric approach \cite{MPT3} in Fig.~\ref{fig:maximal invariant set for double integrator}. The geometric approach treats parameter as unknown bounded signals, and the control input is free from any state-feedback structure. Not surprisingly, the set $\bm{\Breve{\mathcal{S}}}$ (volume $21.7907$) computed using the proposed approach was found to be larger than the maximal RCI set $\bm{\Omega}_{\infty}$ (volume $19.3703$). Moreover, the overall representational complexity of the set $\bm{\Breve{\mathcal{S}}}$ is just $8$, which is exactly half the complexity of the set $\bm{\Omega}_{\infty}$, this further  demonstrates the benefits of using PD-RCI sets and PDCL in the LPV setting. 
We also show the set $\bm{\Breve{\mathcal{S}}}_{\gamma}$ in Fig.~\ref{fig:maximal invariant set for double integrator}, which satisfy performance constraints $\sum_{t=0}^{\infty}x(t)^T Q_x x(t)+u(t)^T Q_u u(t)\leq\gamma$, for all $x$ within the set. Where $Q_x=I$, $Q_u=0.1$ and $\gamma=10$. 
Lastly, to demonstrate the main advantage of the presented algorithm, we perform an analysis in which the RCI sets are computed by changing the bound $\theta_{max}$ on the parameter $\theta$. The volume of the computed set $\bm{\Breve{\mathcal{S}}}$ is plotted against parameter bound in Fig.~\ref{fig:parameter_var}. As expected from the theory, the volume decreases with an increase in the value of $\theta_{max}$. Nonetheless, it is interesting to observe that the proposed method is able to compute the RCI sets even for a large bound on the scheduling parameter. We remark that the geometric approach \cite{MPT3}, failed to generate any RCI set for $\theta_{max}\geq 0.4$
\subsection{Nonlinear System}
\begin{figure}
    \centering
    \includegraphics[width =5.5cm, height=4cm]{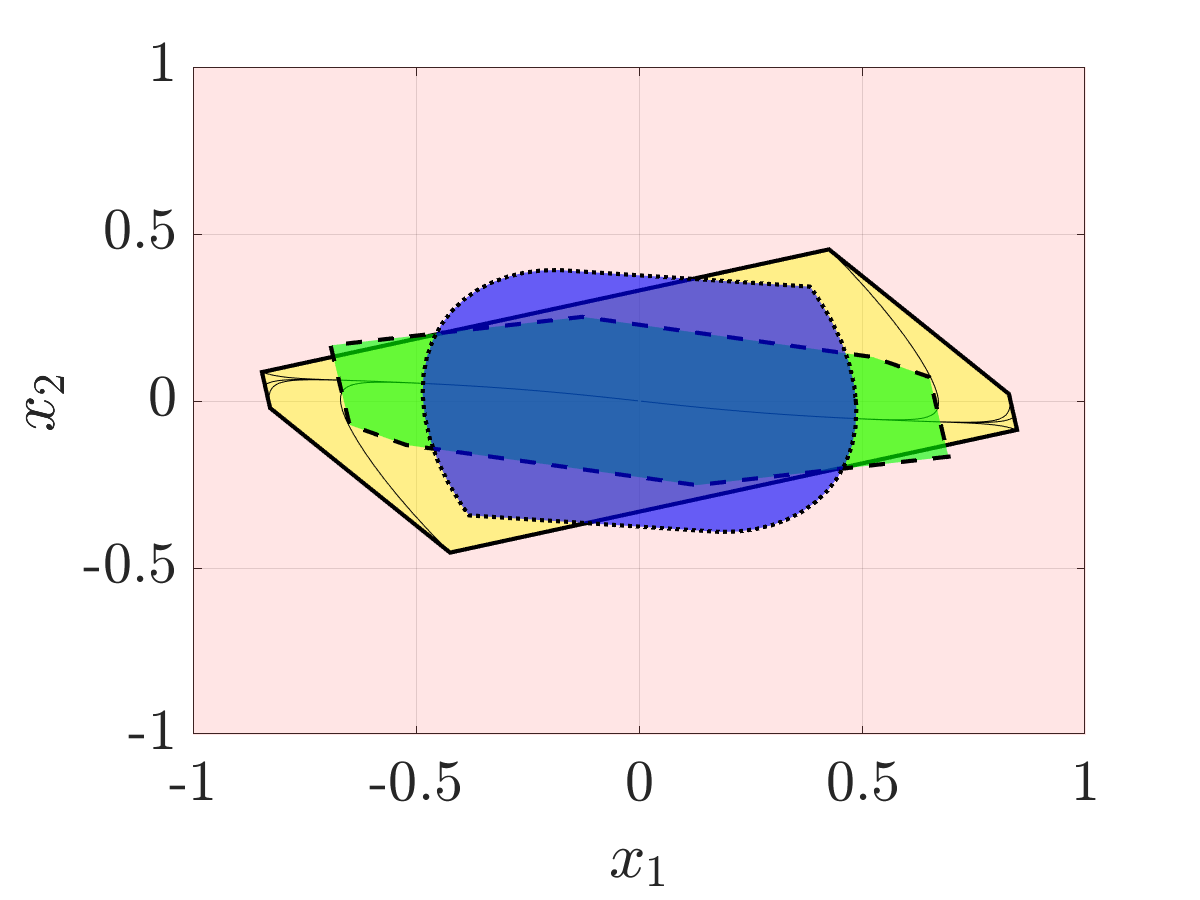}
    \caption{Admissible set $\bm{\mathcal{X}}$ (red), RCI set $\bm{\Breve{\mathcal{S}}}$ (yellow; solid), RCI set using \cite{ag19a} (green; dashed) and RPI set (blue; dotted)  for the Van der Pol oscillator system.}
    \label{fig:vander pol oscillator}
\end{figure}

One important application of the proposed approach is to compute RCI sets for nonlinear systems. For this purpose, we consider the  controlled Van der Pol oscillator system in \cite{jh17}: \vspace{-0.2cm}
\begin{equation}\label{eq:nonlinear example}
    \dot{x}_1 = x_2,\, \dot{x}_2 = -x_1 + \mu(1-x_1^2)x_2+u,\vspace{-0.2cm}
\end{equation}
where $\mu = 2$. The system should satisfy the input constraints $|u|\leq1$ and state constraints $|x_1|\leq1$, $|x_2|\leq1$. For computation and simulation purpose we discretize the system using Eulers method with sampling time $0.1$ units. Further, we rewrite the system in the quasi-LPV form \eqref{eq:affine lpv representation} with scheduling parameters  $\xi_1=(2-\mu(1-x_1^2))/2$ and $\xi_2=\mu(1-x_1^2)/2$. %by selecting \vspace{-0.2cm}
%\begin{equation}
%\begin{bmatrix}
%\begin{array}{c|c} 
%A^1&B^1\\\hline
%A^2&B^2
%\end{array}
%\end{bmatrix}
%=
%\begin{bmatrix}
%\begin{array}{cc|c}
%\;\;\,1.0 & \;\;\,0.1 & 0 \\ 
%     -0.1 & \;\;\,1.0 & 0.1\\\hline
%\;\;\,1.0 & \;\;\,0.1 & 0 \\ 
%     -0.1 & \;\;\,1.2 & 0.1
%\end{array}
%\end{bmatrix},\vspace{-0.2cm}
%\end{equation}
%\begin{align}
%    A^1 = \begin{bmatrix}
%    \;\;\,1.0&\;\;\,0.1\\-0.1&\;\;\,1.0
%    \end{bmatrix}, A^2=\begin{bmatrix}
%    \;\;\,1.0&\;\;\,0.1\\-0.1&\;\;\,1.2
%    \end{bmatrix},\\
%    B^1 = B^2 = \begin{bmatrix}
%    0.0\\0.1
%    \end{bmatrix},
%\end{align}
 Using the proposed approach we compute the matrix variables defining the RCI set and the invariance inducing controller for the nonlinear system which are given as
\vspace{-0.2cm}
\begin{align*}%\nonumber
   [P_1|P_2] &= \begin{bmatrix}
   \begin{array}{cc|cc}
-0.5066  & -0.1205 & -0.5066 &  -0.1358\\
-0.4349  & -0.0135 & -0.4367 & -0.0134\\
\;\;\,0.4238 &  -0.2686 & \;\;\, 0.4237 &  -0.3173\\
\;\;\,0.5280  & \;\;\, 0.0385 & \;\;\,0.5280 &   \;\;\,0.0385
\end{array}
\end{bmatrix},\\
\begin{bmatrix}\!
\begin{array}{c|c}
W&\!\begin{array}{c} K^1\\ \hline K^2\end{array} \\ 
\end{array}\!\!\end{bmatrix}&=
\begin{bmatrix}\!\!
\begin{array}{c|c}
\begin{array}{cc}
\;\;\,0.4409&    \;\;\,0.0136\\
   -0.0127  &  \;\;\,0.1090
\end{array}&\! \begin{array}{cc} \;\;\,0.8341 &  -2.3111\\\hline\;\;\,0.8727 &  -3.0114\end{array}
\end{array} \!\!
\end{bmatrix}.
%\left [ \begin{array}{c}
%W\\\hline
%K_1\\\hline
%K_2\\
%\end{array} \right ]& = \left [ %\begin{array}{cc}
%\;\;\,0.4409&    \;\;\,0.0136\\
%   -0.0127  &  \;\;\,0.1090\\\hline
%\;\;\,0.8341 &  -2.3111\\\hline
%\;\;\,0.8727 &  -3.0114
%\end{array} \right ]
\vspace{-0.2cm}
\end{align*}
Since the scheduling parameters $\xi_1,\xi_2$ are state dependent, in accordance with Section~\ref{sec:quasi LPV}, we compute RCI set $\bm{\Breve{\mathcal{S}}}$ \eqref{eq:RCI set with finite intersection}, shown in Fig.~\ref{fig:vander pol oscillator}. 
%Furthermore, due to state dependence of the parameters, we can rewrite the PDCL controller as a nonlinear state-feedback controller $u=\mathcal{K}(\xi)x = x_2(0.7003x_1^2-3.011)-x_1(0.03855x_1^2-0.8727)$. 
The closed-loop trajectories from all the vertices of the set $\bm{\Breve{\mathcal{S}}}$ are also shown in Fig.~\ref{fig:vander pol oscillator}. For comparison, we compute an RCI set (of a representational complexity same as $\bm{\Breve{\mathcal{S}}}$) using the method presented in \cite{ag19a}, which assumes the invariance inducing controller to be linear state-feedback. We show the computed set in Fig.~\ref{fig:vander pol oscillator} with green color. It can be seen that this set is smaller than the one generated by the proposed algorithm presented in this paper. The geometric approach \cite{MPT3} for computing maximal RCI set did not converge even after $24 hrs$, so instead, we show a robust positive invariant (RPI) set corresponding to an LQR controller for nominal system and tuning matrices $Q=I$ and $R=1$. The representational complexity of the RPI set is $50$.  Clearly, the proposed algorithm is more advantageous here since it can generate visibly larger RCI sets of low complexity.
%Clearly, the presented approach computes a set which is visibly larger than the one presented in \cite{hc98,mc03}. In the above example we selected the inclusion polytope heuristically, in practice a systematic approach should be used else we may obtain a conservative PDRCI sets, see \cite{mc03}.
\section{Conclusion}\label{sec:conclusions}
The paper presented a novel iterative algorithm to compute a PD-RCI set and PD-invariance inducing control law for LPV systems. At each iteration of the algorithm, an SDP is solved to obtain a larger PD-RCI set successively until convergence. In the SDP, we introduced the invariance conditions, system constraints and performance constraints as LMIs, which were constructed using   Finslers's lemma and zeroth order Polya's relaxation. Besides, we also presented a new approach for volume maximization of polytopes based on Monte-Carlo principles. It was shown that a larger invariant set could be obtained by exploiting the knowledge of parameters in the invariant set description as well as in the controller design. We assumed candidate RCI set to be 0-symmetric. This is a reasonable assumption if the system is linear and the constraints are 0-symmetric. In other cases, this assumption would be potentially conservative. Thus, a natural extension of this work could be to devise a similar algorithm for computing non-symmetrical RCI set. %To further reduce conservatism, another possible extension is to include rate constraints on the scheduling parameter variation.
\begin{ack}\vspace{-0.2cm}
The  authors  thank Dr. Hakan K\"{o}ro\u{g}lu for his valuable inputs.\vspace{-0.2cm}
\end{ack}

  \appendix

 \section{Volume maximization using Monte-Carlo integration}\label{sec:volume maximization}

 Based on the theory of Monte-Carlo integration \cite{robert2004}, we present an approach which can be used to find a desirably large   polytope of a predefined maximum complexity, enclosed within some known set. %The resulting optimization problem is non-convex. However, we show that by introducing appropriate modifications to the objective function, the overall problem can be posed as a tractable semi-definite program.

 Let $\bm{\mathcal{C}}$ be a set defined as,
 \begin{equation*}
     \bm{\mathcal{C}}=\{x\in \mathbb{R}^{n_x}| Px\leq \mathbf{1}\},
 \end{equation*}
  where $P\in \mathbb{R}^{n_p \times n_x}$. %and $\mathrm{rank}(\bar{P}) = n_x$ for the polytopic set $\bm{\mathcal{C}}$ to be full-dimensional. 
% The set can be equivalently represented as,
% \begin{equation*}
%    \bm{\mathcal{C}}=\{x\in \mathbb{R}^{n}| \  Px\leq %\end{equation*}
%with $P = [\bar{P}^{T} -\bar{P}^{T}]^{T} \in \mathbb{R}^{2n_p \times n_x}$. 
%and  $\mathrm{rank}(P) = n_x$.  
%For the polytopic set $ \bm{\mathcal{C}}$ to be full-dimensional, it is required that $P$ has a full column rank, \emph{i.e.}, $\mathrm{rank}(P) = n_x$.
 % Let $\bm{\mathfrak{B}}$ be a \emph{known} outer bounding set which contains $\bm{\mathcal{C}}$. 
 We consider the following volume maximization problem,
 \begin{equation}\label{eq:max_vol}
 \sup_{\bm{\mathcal{C}}} \int_{\bm{\mathcal{C}}} dx \quad \mathrm{s.t.} \ \ \bm{\mathcal{C}} \subseteq  \bm{\mathcal{X}},
% %\bm{\mathfrak{B}},
 \end{equation} 
 where $\bm{\mathcal{X}}$ is a given bounded set, not necessarily a polytope.  
%where $\bm{\mathcal{C}}$ is a candidate polytopic set of given complexity $n_p$ to be computed and $\bm{\mathfrak{B}}$ is a known outer bounding set. %Typically, determining the exact volume $\int_{\poly} dx$ of a polytope $\poly$ is  computationally challenging \cite{bueler2000poly,dyer2000vol}. In our case, the problem is even more difficult as $\poly$ itself is \emph{not} known. To this end, 
 We assume that the set containment constraints $\bm{\mathcal{C}} \subseteq  \bm{\mathcal{X}}$ are already available, and %(as well as other constraints on $P$)
 are formulated as some finite-number of convex constraints  (e.g., LMIs). In this section, we focus on the cost function of \eqref{eq:max_vol}, which characterizes the volume of the polytopic set $\poly$. Typically, determining the exact volume $\int_{\poly} dx$ of a polytope $\poly$ is  computationally challenging \cite{bueler2000poly,dyer2000vol}. In our case, the problem is even more difficult as $\poly$ itself is \emph{not} known. To this end, 
 a procedure based on Monte-Carlo methods~\cite{robert2004} is formulated, in order to approximate the cost in \eqref{eq:max_vol}. %and solve problem \eqref{eq:max_vol}. 

%Since $\bm{\mathfrak{B}}$ is known, 
%we can generate $N$ uniformly distributed random samples $\widetilde{\bm{\mathcal{X}}} = \{\widetilde{x}_j
%\}_{j=1}^{N}$. 
  Let $\bm{\mathfrak{B}}$ be a \emph{known} outer bounding box which contains the given set $\bm{\mathcal{X}}$. 
 We generate $N$ independent random samples $\widetilde{\bm{\mathcal{X}}} = \{\widetilde{x}_j\}_{j=1}^{N}$, which are uniformly distributed in the given outer-bounding box $\bm{\mathfrak{B}}$.

 According to Monte-Carlo integration technique, the volume of the set $\bm{\mathcal{C}}$ is approximated as,
 \begin{equation}\label{eq:vol_approx}
 \int_{\bm{\mathcal{C}}} dx \approx \frac{1}{N} \mathsf{vol}(\bm{\mathfrak{B}}) \sum_{\widetilde{x}_j \in \widetilde{\bm{\mathcal{X}}}} \indicator{\bm{\mathcal{C}}}{\widetilde{x}_j},
 \end{equation} 
 where $\mathsf{vol}(\bm{\mathfrak{B}})$ denotes the volume of the box $\bm{\mathfrak{B}}$, and $\indicator{\bm{\mathcal{C}}}{\widetilde{x}_j}$ is the indicator function of the set $\bm{\mathcal{C}}$ defined as,
 \begin{equation}
 \indicator{\bm{\mathcal{C}}}{\widetilde{x}_j} =
 \begin{cases}
 1, \ \mathrm{if} \ \ \widetilde{x}_j \in \bm{\mathcal{C}}, \\
 0, \  \mathrm{if} \ \ \widetilde{x}_j \notin \bm{\mathcal{C}}. 
 \end{cases}
 \end{equation}

 \begin{remark2}
 	From the theory of Monte Carlo
 	integration~\cite{robert2004}, the following limit holds with probability (w.p) $1$:
 	\begin{equation*}
 	\lim\limits_{N \rightarrow \infty} \frac{1}{N} \mathsf{vol}(\bm{\mathfrak{B}}) \sum_{\widetilde{x}_j \in \widetilde{\bm{\mathcal{X}}}} \indicator{\bm{\mathcal{C}}}{\widetilde{x}_j} = \int_{\poly} dx , \quad \mathrm{w.p.} 1.
 	\end{equation*}
 \end{remark2}

 By using \eqref{eq:vol_approx}, the volume maximization problem \eqref{eq:max_vol} is approximated as follows,
 \begin{equation}\label{eq:max_vol_approx}
% %\sup_{\bm{\mathcal{C}}}
 \max_{\mathcal{C}} \sum_{\widetilde{x}_j \in \widetilde{\bm{\mathcal{X}}}} \indicator{\bm{\mathcal{C}}}{\widetilde{x}_j} \quad \mathrm{s.t.} \ \ %\bm{\mathcal{C}} \subseteq \bm{\mathfrak{B}}.
 \bm{\mathcal{C}} \subseteq \bm{\mathcal{X}}.
 \end{equation}

Note that the cost function in \eqref{eq:max_vol_approx} is the sum of indicator functions $\indicator{\bm{\mathcal{C}}}{\widetilde{x}_j}$, which is non-convex and discontinuous. We next introduce an approximation of the cost function in order to solve \eqref{eq:max_vol_approx} in a tractable manner. %The cost in \eqref{eq:max_vol_approx} assigns maximum value of $1$, for each $\widetilde{x}_j \in \poly$. %to the sample points $\widetilde{x}_j$, which are inside the set $\bm{\mathcal{C}}$. Based on this rationale,
 We approximate the discontinuous cost in  \eqref{eq:max_vol_approx} with a continuous-concave function such that the sample points which are contained in $\bm{\mathcal{C}}$ get the maximum cost, while the value of the cost for all $\widetilde{x}_j \notin \poly$, decreases uniformly. %This, in turn, maximizing  proportional to the volume of the set $\bm{\mathcal{C}}$. 

 For the sake of convenience, without loss of generality we modify the definition of indicator functions  in the cost \eqref{eq:max_vol_approx} as follows
 \begin{equation}\label{eq:indicator}
 \indicator{\bm{\mathcal{C}}}{\widetilde{x}_j} =
 \begin{cases}
 \;\;\,0, \ \ \ \mathrm{if}\, P \widetilde{x}_j - \mathbf{1} \leq 0, \\
 -1, \ \ \mathrm{otherwise}.   
 \end{cases}
 \end{equation}
 Note that, this modification does not change the optimal solution of problem \eqref{eq:max_vol_approx}.

subsubsection{Approximation of the indicator functions}
%Consider the discontinuous objective function in \eqref{eq:max_vol_approx_b}, given by the sum of indicator functions  
%defined as,
%\begin{equation*}
%\indicator{\bm{\mathcal{C}}_i}{\widetilde{x}_j} =
%\begin{cases}
%1, \ \mathrm{if} \ \  \widetilde{x}_j \in %\bm{\mathcal{C}}_i,\\
%0, \  \mathrm{if} \ \ \widetilde{x}_j \notin %\bm{\mathcal{C}}_i.
%\end{cases}
%\end{equation*}
% Following the rationale stated above (also see \cite{piga2012cdc,PigaSetmem19}), we present the following continuous piecewise-linear function $\Tc{\bm{\mathcal{C}}}{\widetilde{x}_j}$ to approximate the indicator functions defined $\indicator{\bm{\mathcal{C}}}{\widetilde{x}_j}$ in \eqref{eq:indicator}, 

 Let us first consider for each individual hyperplane $\poly_i \triangleq \{x: e^{T}_{i}(Px-  \mathbf{1}) \leq 0 \}$ of the set $\poly$, the following cost $\Tcn{\bm{\mathcal{C}}_i}{\widetilde{x}_j}$,
 \begin{align}\label{eq:Ta}
 \Tcn{\bm{\mathcal{C}}_i}{\widetilde{x}_j} =
 \begin{cases}
 0   &\mathrm{if} \ \  e^{T}_{i}(P\widetilde{x}_j -  \mathbf{1}) \leq 0,\\
 -e^{T}_{i}(P \widetilde{x}_j- \bm{1})  &\mathrm{if} \ \ e^{T}_{i}(P\widetilde{x}_j -  \mathbf{1}) > 0,
 \end{cases}
 \end{align}
 which is a piecewise linear concave approximation of the indicator functions $\indicator{\poly_i}{\widetilde{x}_j}$ defined for the $i$-th hyperplance of $\poly$.
 The plot of the indicator function $\indicator{\poly_i}{\widetilde{x}_j}$ and its approximation  $\Tcn{\bm{\mathcal{C}}_i}{\widetilde{x}_j}$ is shown in Fig.\ref{fig:T}. The idea of approximating non-convex indicator  function $\indicator{\bm{\mathcal{C}}_i}{\widetilde{x}_j}$ with $\Tcn{\bm{\mathcal{C}}_i}{\widetilde{x}_j}$ is similar to the relaxation of $l_{0}$-quasi-norm with $l_{1}$-norm as introduced in \cite{ben2016,PigaSetmem19} for computing outer-approximating polytopes of non-convex semialgebraic sets. 
%Such approximation is used in the computation of the sparsest solution, commonly employed in the field of compressive sensing and in identification for model order selection. The interested reader is referred to~\cite{tibshirani1996regression,boyd2004convex,piga2012cdc} for more details. 

% \begin{align}\label{eq:T}
% \Tc{\bm{\mathcal{C}}_i}{\widetilde{x}_j} =
% \begin{cases}
% 1   &\mathrm{if}\,  e^{T}_{i}(P\widetilde{x}_j - \mathbf{1}) \leq 0,\\
% -(e^{T}_{i}(P \widetilde{x}_j- \mathbf{1}))+1  &\mathrm{if}\, e^{T}_{i}(P\widetilde{x}_j -  \mathbf{1}) > 0.
% \end{cases}
% \end{align}
% \begin{align}\label{eq:T}
%  \Tc{\bm{\mathcal{C}}}{\widetilde{x}_j} =
%  \begin{cases}
%  0   &\mathrm{if}\, \widetilde{x}_j \in \bm{\mathcal{C}}, \\
%  %-\left \|P \widetilde{x}_j- \bm{1}  \right \|_{\infty}   &\mathrm{if}\, \widetilde{x}_j \notin \bm{\mathcal{C}}, 
%  - \left( \max \limits_{i=1,\ldots,n_p}\e^{T}_{i}(P \widetilde{x}_j- \bm{1}) \right) &\mathrm{if}\, \widetilde{x}_j \notin \bm{\mathcal{C}}
%  \end{cases}
%  \end{align}
%  \emph{i.e.,}

%Following the rationale stated above, 
 We now extend the idea of approximating the indicator functions defined for a single hyperplane  $\poly_{i}$, to approximate the indicator function defined over the entire polytopic set $\poly$. In particular, we introduce the following concave function $\Tc{\bm{\mathcal{C}}}{\widetilde{x}_j}$ to approximate $\indicator{\bm{\mathcal{C}}}{\widetilde{x}_j}$ defined in \eqref{eq:indicator}, %$\Tc{\bm{\mathcal{C}}}{\widetilde{x}_j}$ for the  polytopic set $\poly$,
  \begin{align}\label{eq:T1}
  \Tc{\bm{\mathcal{C}}}{\widetilde{x}_j} =
  \begin{cases}
  0  & \!\! \! \mathrm{if}\, P \widetilde{x}_j \!-\! \mathbf{1} \leq 0, \\
  - \left( \max \limits_{i=1,\ldots,n_p}e^{T}_{i}(P \widetilde{x}_j- \bm{1}) \right) &\!\! \mathrm{otherwise}.
  \end{cases}
  \end{align}
 
 Note that, for the points $\widetilde{x}_j \notin \bm{\mathcal{C}}$, the cost $\Tc{\bm{\mathcal{C}}}{\widetilde{x}_j}$ is always negative and decays uniformly in all the directions away from $\bm{\mathcal{C}}$. %\MMr{of the maximum over the $n_p$ linear cost $e^{T}_{i}(P \widetilde{x}_j- \bm{1})$ of the of individual hyperplanes.}%We remark that, as the translation $\Tc{\bm{\mathcal{C}}}{\widetilde{x}_j}$ preserves the concavity of $\Tc{\bm{\mathcal{C}}}{\widetilde{x}_j}$ as well as the argument at which the maximum is attained, we consider the following translated function for the sake of convenience,

\begin{figure}[tb!]
	\centering
	\begin{tikzpicture}
	\begin{axis}[axis x line=middle, axis y line=middle,
		ymin=-1.2, ymax=0.2,  
		xmin=-2.3, xmax=2.3,  xlabel=$e^{T}_{i}(P \widetilde{x}_j -  \mathbf{1})$,
		domain=-3.3:3.3,xticklabels={},
		yticklabels={}] 
	\addplot[domain=-2:0,blue, very thick] {0};
	\addplot[domain=0:2,  blue, very thick] {-1};
	\addplot[domain=0:2,red, very thick] {-x};
%	\coordinate (two) at (axis cs:2.25,0.5) {f};
 \node[] at (axis cs: 1.2,-0.5) {$\Tcn{\bm{\mathcal{C}}_i}{\widetilde{x}_j}$};
  \node[] at (axis cs: -1.8,0.1) {$\indicator{\bm{\mathcal{C}}_i}{\widetilde{x}_j}$};
  \node[] at (axis cs: -0.25,-1) {$-1$};
  \node[] at (axis cs: -0.14,0.05) {$0$};
%	\draw[dotted] (axis cs:6,6) -- (axis cs:6,-5);
%	\addplot[holdot] coordinates{(0,0)(4,4)(6,-5)};
%	\addplot[soldot] coordinates{(4,16)(6,6)(10,-5)};
\end{axis}
\end{tikzpicture}
\caption{Indicator function $\indicator{\bm{\mathcal{C}}_i}{\widetilde{x}_j}$ (blue) and its concave approximation $\Tcn{\bm{\mathcal{C}}_i}{\widetilde{x}_j}$ (red) for the $i$-th hyperplane. When $e^{T}_{i}(P \widetilde{x}_j -  \mathbf{1}) < 0$, $\indicator{\bm{\mathcal{C}}_i}{\widetilde{x}_j}$ and $\Tcn{\bm{\mathcal{C}}_i}{\widetilde{x}_j}$ are overlapped and both are $0$.}
\label{fig:T}
\end{figure}
 Based on the approximation $\Tc{\bm{\mathcal{C}}}{\widetilde{x}_j}$ in \eqref{eq:T1} of the indicator functions $\indicator{\bm{\mathcal{C}}}{\widetilde{x}_j}$, the problem \eqref{eq:max_vol_approx} is relaxed as follows,
% % \begin{equation}\label{eq:max_vol_approx_T}
% % \max_{P}  \sum_{\widetilde{x}_j \in \widetilde{\bm{\mathcal{X}}}}  \sum_{i=1}^{n_p} \Tc{\bm{\mathcal{C}}_i}{\widetilde{x}_j} \quad \mathrm{s.t.} \ \ \bm{\mathcal{C}} \subseteq \bm{\mathfrak{B}}, \ \mathrm{rank}(P)=n_x.
% % \end{equation}
 \begin{equation}\label{eq:max_vol_approx_T}
 \max_{\mathcal{C}}  \sum_{\widetilde{x}_j \in \widetilde{\bm{\mathcal{X}}}}   \Tc{\bm{\mathcal{C}}}{\widetilde{x}_j} \quad \mathrm{s.t.} \ \ \bm{\mathcal{C}} \subseteq \bm{\mathcal{X}}
 \end{equation}

 Thus, by solving  the constraint optimization problem \eqref{eq:max_vol_approx_T}, we try to find the matrix $P$ (defining the polytope $\poly$), which maximizes the number of points $\widetilde{x}_j$ inside the set $\poly$, in turn, maximizing its volume, while respecting the constraint $\poly \subseteq \bm{\mathcal{X}}$. 
%The rank constraint $\mathrm{rank}(P) = n_x$ ensures that the polytope  $\bm{\mathcal{C}}$ is full dimensional. 

%\MM{\begin{rem}\label{rem:choice_of_sample}
%With the choice of cost function $\Tc{\bm{\mathcal{C}}_i}{\widetilde{x}_j}$ in \eqref{eq:Ta}, we aim at selecting the hyperplanes of the polytope $\bm{\mathcal{C}}$, such that they are ``pushed" closer to the boundaries of the outer-bounding box $\bm{\mathcal{B}}$. This is due to the fact that,  the value of the cost $\Tc{\bm{\mathcal{C}}_i}{\widetilde{x}_j}$ decreases linearly for the sample points $\widetilde{x}_j$   which lie outside the region defined by the $i$-th hyperplane, i.e., sample points which are farther away from the hyperplane are penalized less. 
%We observe that, with this choice of the cost function, it is sufficient to select the vertices of the outer-bounding box $\bm{\mathcal{B}}$ as the sample points $\{\widetilde{x}_j\}_{j=1}^{N_{\sigma}}$. This significantly reduces the computation cost to solve the optimization problem~\eqref{eq:max_vol_approx_T}. Thus, in this paper, we have considered the sample points to be the vertices  of $\bm{\mathcal{B}}$, instead of uniformly distributed samples.      
%\end{rem}
% }
 \begin{remark2}\label{rem:choice_of_sample}
 With the choice of cost function $\Tc{\bm{\mathcal{C}}}{\widetilde{x}_j}$ in \eqref{eq:T1}, we aim at selecting the matrix $P$ of the polytope $\bm{\mathcal{C}}$, such that maximum number of points lie in the set, i.e.,  $\widetilde{x}_j\in \bm{\mathcal{C}}$. This is due to the fact that,  the value of the cost $\Tc{\bm{\mathcal{C}}}{\widetilde{x}_j}$ decreases linearly for the sample points $\widetilde{x}_j$ which lie outside the set $\bm{\mathcal{C}}$. 
 We observe that, with this choice of the cost function, it is sufficient to select the %vertices of the 
% %outer-bounding box $\bm{\mathcal{B}}$ as the sample points $\{\widetilde{x}_j\}_{j=1}^{N_{\sigma}}$. 
  sample points $\{\widetilde{x}_j\}_{j=1}^{N_{\sigma}}$ which lie on the boundary of the known outer-bounding box $\bm{\mathfrak{B}}$. 
 This significantly reduces the computation cost to solve the optimization problem~\eqref{eq:max_vol_approx_T}. Thus, in this paper, we have considered only the sample points which are on the boundary of  $\bm{\mathfrak{B}}$, instead of uniformly distributed samples.      
 \end{remark2}

 Finally, the cost function in \eqref{eq:max_vol_approx_T} can be seen as a sum of concave functions, which can be equivalently expressed as following convex \emph{minimization} problem,
% \begin{subequations}
% \begin{align}\label{eq:max_vol_approx_LP}
% &\min_{\MM{\GG,  \bb, t_{ij}}}  \sum_{j=1}^{N}  \sum_{i=1}^{n_p} t_{ij}\\
% &\mathrm{s.t.} \ \  \GG^{i}p_j - \bb^{i} \leq t_{i,j} \ \ \forall i=1,\ldots, n_p, \ p_j \in %\mathcal{P} \\
% & \quad 0 \leq t_{ij}  \ \ \forall i=1,\ldots, n_p, \ p_j \in \mathcal{P}  \\
% & \quad  \poly^{k+1} \subseteq \mathcal{X}, \ \MM{\mathcal{L}\{{\poly^{k+1}, \poly^{k}}\} \leq 0}.
% \end{align} 
% \end{subequations}

% \begin{align}\label{eq:max_vol_approx_LP}
% \begin{array}{cc}
%   \min   &  \sum_{j=1}^{N}\bm{1}^T \sigma_j\\
%   {P,\sigma_j}\\
%   \text{s.t.}  & \sigma_j \geq 0,  \ \ \forall j=1,\ldots, N\\
%  & P\widetilde{x}_j-\mathbf{1}\leq\sigma_j,  \ \ \forall j=1,\ldots, N\\
%  & \bm{\mathcal{C}} \subseteq \bm{\mathfrak{B}}, \ \mathrm{rank}(P)=n_x.
% \end{array}
% \end{align}
% where $\sigma_j \in \mathbb{R}^{n_p}$. 

 \begin{align}\label{eq:max_vol_approx_LP}
 \begin{array}{cc}
   \min   &  \sum_{j=1}^{N} \sigma_j\\
   {P,\sigma_j}\\
   \text{s.t.}  & \sigma_j \geq 0,  \ \ \forall j=1,\ldots, N\\
  & P\widetilde{x}_j - \bm{1}\leq\sigma_j \bm{1},  \ \ \forall j=1,\ldots, N\\
  & \bm{\mathcal{C}} \subseteq \bm{\mathcal{X}}, \ 
 \end{array}
 \end{align}
 where $\sigma_j \in \mathbb{R}$. Thus, the final volume maximization problem consist of a linear cost and constraints, alongwith a set containment constraint.

%We remark that, for our problem under consideration, 
%the set containment constraints of the form 
%$\bm{\mathcal{C}} \subseteq \bm{\mathcal{X}}$ 
%(e.g., invariance conditions and state constraints) have been formulated as LMIs. These LMI conditions also ensure that the rank condition  $\mathrm{rank}(P)=n_x$ is satisfied and thus, $P$ has a full column rank for the polytopic set $\bm{\mathcal{C}}$ to be full dimensional. Thus, problem \eqref{eq:max_vol_approx_LP} is a convex semi-definite program.  

%{\color{red} mention about the assumption that the condition $\bm{\mathcal{C}} \subseteq \bm{\mathfrak{B}}$ should also assure that $P$ has full colum rank.}
%\bibliographystyle{plain}        % Include this if you use bibtex

\bibliographystyle{agsm}
\bibliography{autosam}        % and a bib file to produce the bibliography

@article{sr10,
author = {Rakovi{\'{c}}, S. V. and Baric, M.},
journal = {IEEE Transactions on Automatic Control},
month = {Jul},
number = {7},
pages = {1599--1614},
title = {{Parameterized Robust Control Invariant Sets for Linear Systems: Theoretical Advances and Computational Remarks}},
volume = {55},
year = {2010}
}

@article{tb10,
author = {Blanco, T. B. and Cannon, M. and {De Moor}, B.},
journal = {International Journal of Control},
month = {July},
number = {7},
pages = {1339--1346},
title = {{On efficient computation of low-complexity controlled invariant sets for uncertain linear systems}},
volume = {83},
year = {2010}
}

@book{fb,
address = {Boston, MA},
author = {Blanchini, F. and Miani, S.},
title = {Set-Theoretic Methods in Control},
publisher = {Birkh{\"a}user},
year = {2015}
}

@article{ag19a,
author={Gupta, A. and Falcone, P.},
journal={IEEE Control System Letter},
title={Full-Complexity Characterization of Control-Invariant Domains for Systems with Uncertain Parameter Dependence},
year={2019},
volume={3},
number = {1},
pages = {19--24}
}

@article{ag19b,
title = "Computation of low-complexity control-invariant sets for systems with uncertain parameter dependence",
author = "Gupta, A and  K{\"{o}}ro{\u{g}}lu, H. and  Falcone, P.",
journal = "Automatica",
volume = "101",
pages = "330 - 337",
year = "2019",
}

@article{mk96,
title = "Robust constrained model predictive control using linear matrix inequalities",
journal = "Automatica",
volume = "32",
number = "10",
pages = "1361 - 1379",
year = "1996",
author = "Kothare, M. and Balakrishnan, V. and  Morari, M.",
}

@inproceedings{yalmip,
address = {Taipei, Taiwan},
author = {L{\"{o}}fberg, J.},
booktitle = {Computer-Aided Control System Design Conference},
title = {YALMIP : A Toolbox for Modeling and Optimization in MATLAB},
year = {2004}
}

@InProceedings {MPT3,
    author={M. Herceg and M. Kvasnica and C. N. Jones and M. Morari},
    title={{Multi-Parametric Toolbox 3.0}},
    booktitle={European Control Conference},
    year={2013},
    address={Z\"urich, Switzerland},
    month={July 17--19},
    pages = {502--510}
}

@article{lc19,
author = {Liu, C. and Tahir, F. and Jaimoukha, I..},
title = {Full-complexity polytopic robust control invariant sets for uncertain linear discrete-time systems},
journal = {International Journal of Robust and Nonlinear Control},
volume = {29},
number = {11},
pages = {3587-3605},
year = {2019}
}

@article{or05,
title = "Stability of polytopes of matrices via affine parameter-dependent Lyapunov functions: Asymptotically exact LMI conditions",
journal = "Linear Algebra and its Applications",
volume = "405",
pages = "209 - 228",
year = "2005",
author = "Oliveira, R. and Peres, P. L. D."
}

@article{ag20,
author = {Gupta, A. and Köroğlu, H. and Falcone, P.},
title = {Computation of robust control invariant sets with predefined complexity for uncertain systems},
journal = {International Journal of Robust and Nonlinear Control},
volume = {31},
number = {5},
pages = {1674-1688},
year = {2021}
}

@book{robert2004,
	author = {Robert, C. and Casella, G. },
	title = {{M}onte {C}arlo statistical
	methods},
	publisher = {Springer Science},
	year = {2004},
}

@book{bueler2000poly,
	author = {Bueler, B. and Enge, A. and Fukuda, K.},
	title = {Exact volume
	computation for polytopes: a practical study},
	publisher = {DMV
	SEMINAR, Springer},
	volume ={29},
	pages = {131–-154},
	year = {2000}
}

@article{dyer2000vol,
	author ={Dyer, M. and Frieze, A.},
	title ={On the complexity of
	computing the volume of a polyhedron},
	journal ={SIAM Journal
	on Computing},
	year = 1988,
	volume =17,
	pages = {967--974},
}

@article{ip07,
author = {Pólik, I. and Terlaky, T.},
title = {A Survey of the S-Lemma},
journal = {SIAM Review},
volume = {49},
number = {3},
pages = {371-418},
year = {2007}
}

@INPROCEEDINGS{PigaSetmem19,
	author={D. {Piga} and A. {Benavoli}},
	booktitle={18th European Control Conference (ECC)}, 
	title={Semialgebraic Outer Approximations for Set-Valued Nonlinear Filtering}, 
	address={Naples, Italy},
    month={July},
	year={2019},
	volume={},
	number={},
	pages={400-405},}

@INPROCEEDINGS{jh17,
author={J. {Hanema} and R. {Tóth} and M. {Lazar}},
booktitle={Conference on Decision and Control}, 
title={Stabilizing non-linear {MPC} using linear parameter-varying representations}, 
year={2017},
volume={},
number={},
pages={3582-3587}}

@article{sm05,
author={S. {Miani} and C. {Savorgnan}},
journal={Conference on Decision and Control}, 
title={MAXIS-G: a software package for computing polyhedral invariant sets for constrained {LPV} systems}, 
year={2005},
volume={},
number={},
pages={7609-7614},}

@article{mf10,
title = "On the computation of convex robust control invariant sets for nonlinear systems",
journal = "Automatica",
volume = "46",
number = "8",
pages = "1334--1338",
year = "2010",
author = "M. Fiacchini and T. Alamo and E.F. Camacho"
}

@ARTICLE{ji17,
author={J. Y. {Ishihara} and H. T. M. {Kussaba} and R. A. {Borges}},
journal={IEEE Transactions on Automatic Control}, 
title={Existence of Continuous or Constant Finsler's Variables for Parameter-Dependent Systems}, 
year={2017},
volume={62},
number={8},
pages={4187-4193},}

@article{jb05,
title = "On the computation of invariant sets for constrained nonlinear systems: An interval arithmetic approach",
journal = "Automatica",
volume = "41",
number = "9",
pages = "1583--1589",
year = "2005",
author = "J.M. Bravo and D. Limon and T. Alamo and E.F. Camacho",
}

@article{jh20,
title = "Heterogeneously parameterized tube model predictive control for {LPV} systems",
journal = "Automatica",
volume = "111",
pages = "108622",
year = "2020",
author = "J. {Hanema} and R. {Tóth} and M. {Lazar}"
}

@article{fb07,
title = "Stability results for linear parameter varying and switching systems",
journal = "Automatica",
volume = "43",
number = "10",
pages = "1817--1823",
year = "2007",
author = "F. Blanchini and S. Miani and C. Savorgnan",
}

@article{ben2016,
title = "A probabilistic interpretation of set-membership filtering: Application to polynomial systems through polytopic bounding",
journal = "Automatica",
volume = "70",
pages = "158 - 172",
year = "2016",
issn = "0005-1098",
doi = "https://doi.org/10.1016/j.automatica.2016.03.021",
author = "Benavoli, A. and  Piga, D.",
}

@article{ag21ARXIV,
    author  = {A. Gupta and M. Mejari and P. Falcone and D. Piga},
    title   = {Computation of Parameter Dependent Robust Invariant Sets for {LPV} Models
  with Guaranteed Performance},
    year    = {2022},
    journal = {arXiv:2009.09778v1},
    volume  = {},
    pages   = {},
    eprint  = {arXiv:2009.09778v1}
}

@ARTICLE{Nguyen15,  
author={H. {Nguyen} and S. {Olaru} and P. {Gutman} and M. {Hovd}},  
journal={IEEE Transactions on Automatic Control},   title={Constrained Control of Uncertain, Time-varying Linear Discrete-Time Systems Subject to Bounded Disturbances},   year={2015},  
volume={60},  
number={3},  
pages={831-836},  
doi={10.1109/TAC.2014.2346872}}

\end{document}